\documentclass[lettersize,journal]{IEEEtran}
\usepackage{amsmath,amsfonts}
\usepackage{array}
\usepackage{textcomp}
\usepackage{stfloats}
\usepackage{url}
\usepackage{verbatim}
\usepackage{graphicx}
\usepackage{subfigure}
\usepackage{cite}
\usepackage{amssymb}
\usepackage{algpseudocode}
\usepackage{algorithmicx,algorithm}
\usepackage{multirow}
\usepackage{amsmath}
\usepackage{amsthm}
\newtheorem{Lemma}{Lemma}
\newtheorem{theorem}{Theorem}
\newtheorem{corollary}{Corollary}
\makeatletter
\renewcommand{\maketag@@@}[1]{\hbox{\m@th\normalsize\normalfont#1}}%
\makeatother
\begin{document}
\allowdisplaybreaks[4]
\begin{sloppypar}

\title{Split Federated Learning for Low-Altitude Wireless Networks: Joint Sensing, Communication, Computation, and Control Co-design}

\author{Xiangwang Hou, \emph{ Member, IEEE,} Xianghe Wang, Jiacheng Wang, Zekai Zhang,  \\ Jun Du, \emph{Senior Member, IEEE}, Jingjing Wang, \emph{Senior Member, IEEE}, and Yong Ren,  \emph{Senior Member, IEEE}

\thanks{Xiangwang Hou, Xianghe Wang, Jun Du and Yong Ren are with the Department of Electronic Engineering, Tsinghua University, Beijing  100084, China (E-mail: xiangwanghou@163.com; wangxh24@mails.tsinghua.edu.cn; jundu@tsinghua.edu.cn; reny@tsinghua.edu.cn.)}
\thanks{Jiacheng  Wang is with the College of Computing and Data Science, Nanyang Technological University, Singapore 639798. (E-mail: jcwang\_cq@foxmail.com)}
\thanks{Z. Zhang is with Tsinghua Shenzhen International Graduate School, Tsinghua University, Shenzhen 518055, China (E-mail: zhangzekai\_2000@163.com.) }
\thanks{Jingjing Wang is with the School of Cyber Science and Technology, Beihang University, Beijing 100191, China. (Email: drwangjj@buaa.edu.cn)}

\thanks{A part of this work was accepted by IEEE ICC 2025.}
}



\maketitle

\begin{abstract}
Unmanned aerial vehicles (UAVs) with integrated sensing, communication, computation and control (ISC3) capabilities have become key enablers of next-generation wireless networks. Federated edge learning (FEL) leverages UAVs as mobile learning agents to collect data, perform local model updates, and contribute to global model aggregation. However, existing UAV-assisted FEL systems face critical challenges, including excessive computational demands, privacy risks, and inefficient communication, primarily due to the requirement for full-model training on resource-constrained UAVs. To deal with aforementioned challenges, we propose Split Federated Learning for UAV-Enabled ISC3 (SFLSC3), a novel framework that integrates split federated learning (SFL) into UAV-assisted FEL. SFLSC3 optimally partitions model training between UAVs and edge servers, significantly reducing UAVs' computational burden while preserving data privacy. We conduct a theoretical analysis of UAV deployment, split point selection, data sensing volume, and client-side aggregation frequency, deriving closed-form upper bounds for the convergence gap. Based on these insights, we conceive a joint optimization problem to minimize the delay required to achieve a target model accuracy. Given the non-convex nature of the problem, we develop a low-complexity algorithm to efficiently determine UAV deployment, split point selection, and communication frequency. Extensive simulations on a target motion recognition task validate the effectiveness of SFLSC3, demonstrating superior convergence  and delay performance compared to baseline methods.
\end{abstract}

\begin{IEEEkeywords}
Federated edge learning, unmanned aerial vehicle (UAV), split federated learning, integrated sensing, communication, computation, and control (ISC3).
\end{IEEEkeywords}

\section{Introduction}\label{se:1}

\IEEEPARstart{T}{he} rapid development of the low-altitude economy (LAE) is driving a wide range of emerging low-altitude applications, including aerial logistics, disaster response, and urban sensing.
As a core carrier of such applications, unmanned aerial vehicles (UAVs) have evolved beyond their traditional roles by leveraging their inherent mobility and onboard intelligence to support complex low-altitude operations, thereby giving rise to low-altitude intelligent systems characterized by integrated sensing, communication, computation, and control (ISC3) \cite{9468714}.  Their mobility and onboard processing capabilities allow them to not only sense and interact with the environment but also support distributed intelligence at the network edge \cite{Zhang2024WhenUM}\cite{10438010}. However, training machine learning (ML) models in such decentralized and dynamic environments presents significant challenges, particularly in terms of data privacy and communication efficiency \cite{wxh}.

Federated edge learning (FEL),  as an efficient distributed learning paradigm that enables collaborative model training across edge devices without requiring raw data exchange \cite{9220170}\cite{10063977}. UAVs, with their ability to collect real-time data and perform local computation, can play a crucial role in FEL by serving as mobile learning agents \cite{ISAC-2306}. Specifically, they can acquire task-relevant data through onboard sensing, perform local training, and iteratively exchange model updates with a central server, refining a global model in a collaborative and efficient manner. This synergy between UAV-enabled data collection and distributed learning paves the way for more adaptive and intelligent wireless networks.

Despite the promising application prospects of UAV-assisted FEL, practical deployment remains challenging due to several key obstacles. First, UAVs are inherently constrained by limited computational power and battery capacity, making it impractical for them to train full ML models with high-dimensional parameters—such as those in classical models like AlexNet \cite{A4} and VGG16 \cite{A5}. The high computational demands and delay associated with full-model training often exceed the operational capabilities of resource-limited UAVs, significantly restricting the feasibility of existing FL approaches \cite{10630700}. Second, in federated learning (FL), the complete model is typically accessible to all involved parties, including UAVs and servers, raising concerns regarding privacy leakage \cite{B3}. Additionally, UAV's deployment affects both data sensing and communication efficiency, which in turn impacts the overall performance of FL systems \cite{ISAC-2306}.

While previous studies have focused on optimizing UAV deployment within traditional FL frameworks, little attention has been given to redesigning the FL architecture to fully capitalize on the unique advantages of UAVs. Split federated learning (SFL) has been proposed to alleviate the computational load by partitioning complex ML models into client-side and server-side sub-models, thereby improving privacy protection and reducing the local computational burden on edge devices \cite{B1}. However, designing an optimal SFL framework tailored for UAV-assisted FEL systems, alongside effective UAV deployment strategies to enhance overall system performance, remains an open research problem that requires further investigation.

\subsection{Related Works}
At the early stages, many studies employed UAVs as communication relays or edge servers to collect data or aggregate models, thereby overcoming communication bottlenecks in FEL. For example, Zhang \textit{et al}. \cite{Zhang2024WhenUM} considered the UAV as a central node to collaboratively perform FL with numerous devices for model training. Ng \textit{et al}. \cite{9292475} employed the UAV as a relay between vehicles and the server to improve communication quality, thereby enhancing FL performance.

With the continuous evolution of UAV sensing and computing capabilities, recent research on UAV-assisted FL has moved beyond merely utilizing UAVs for flexible communication coverage. Instead, there is increasing emphasis on leveraging their onboard sensing and computation capabilities to enhance FL performance. Tang \textit{et al}. \cite{ISAC-2306} proposed a UAV-assisted FL framework that quantifies the influence of UAV's deployment strategy on the quality of sensed data and further investigates the impact of UAV's placement and resource allocation on learning performance. As aforementioned, despite these advancements, the practical deployment of UAV-assisted FEL remains challenging due to several critical issues. First, training a complete ML model with high-dimensional parameters imposes excessive computational demands on UAVs. Given their constrained onboard resources, including limited processing power and battery capacity, UAVs often struggle to perform full model training, significantly limiting the feasibility of existing approaches. Second, conventional FL requires all participating devices, including UAVs and servers, to access and update the full model, posing inherent privacy risks.

In this context, the integration of SFL into UAV-assisted FEL systems presents a promising solution. Although no existing work has explored the integration of SFL into UAV-assisted FEL systems, recent studies have investigated fundamental SFL principles, key phenomena, and optimization strategies. A crucial factor is the selection of the split point, which significantly influences the computational load on the client and server, communication overhead, and model convergence speed. Previous research has examined the impact of split point selection. For instance, in \cite{B6,B7}, the authors developed regression models based on the delay and energy consumption of different split points to determine the optimal division. Wu \textit{et al}. \cite{B9} considered device heterogeneity and network dynamics, optimizing split point selection to minimize training costs. Lin \textit{et al}. \cite{B10} addressed the adaptive selection of split points in multi-device scenarios. Hou \textit{et al}. \cite{HouSFL} proposed the SFL-Integrated Sensing, Computation, and Communication (ISCC) framework, integrating model splitting technology into FL-based UAV ISCC systems, and further incorporated semi-supervised learning into the framework to address the issue of sparse labeled data in UAV networks.

Furthermore, the frequency of client aggregation plays a critical role in model convergence and communication overhead. Typically, a higher aggregation frequency accelerates model convergence but also increases communication costs \cite{B11}. In \cite{B10} and \cite{B12}, the authors explored the joint optimization of split point selection and aggregation frequency for balancing training accuracy and communication costs. Some studies have also analyzed the theoretical relationship between communication rounds and convergence speed, aiming to minimize overall training costs (in terms of delay or energy consumption) while satisfying desired convergence performance constraints \cite{B13,B14,B15}. 

Inspired by these studies, this work aims to introduce the concept of split federated learning into UAV-Enabled ISC3, taking UAV-driven collaborative intelligence to a new level.

\subsection{Contributions}
This paper proposes a novel split federated learning framework for UAV-enabled integrated sensing, communication, computation, and control  (SFLSC3). The main contributions are as follows:

\begin{itemize}
    \item  To the best of our knowledge, this is the first work to incorporate SFL into UAV-assisted ISC3 system. We theoretically analyze the impact of UAV deployment, split point selection, data sensing volume, and client-side aggregation frequency on the convergence of SFLSC3 and derive closed-form upper bounds for the convergence gap.
    
    \item  Building upon the convergence analysis, we formulate a joint optimization problem to minimize the delay of the UAV-assisted ISC3 system while ensuring a target model accuracy. The optimization jointly considers split point selection, aggregation frequency, UAV deployment strategy, and data sensing volume.
    
    \item Given the non-convex nature of the formulated problem, we design a low-complexity algorithm to efficiently obtain a near-optimal solution. Furthermore, we conduct extensive experiments on a target motion recognition task using a high-fidelity wireless perception simulator. The results demonstrate that SFLSC3 achieves superior convergence performance and latency performance compared to baseline methods.
\end{itemize}

The rest of this paper is organized as follows. Section II introduces the system model, while section III presents the convergence analysis of SFLSC3. Section IV formulates the optimization problem and provides the solution algorithm. Subsequently, the simulation results are presented in Section V, and the conclusion is provided in Section VI.

\section{System Model}

\begin{figure}[t]
    \centering
\includegraphics[width=0.5\textwidth]{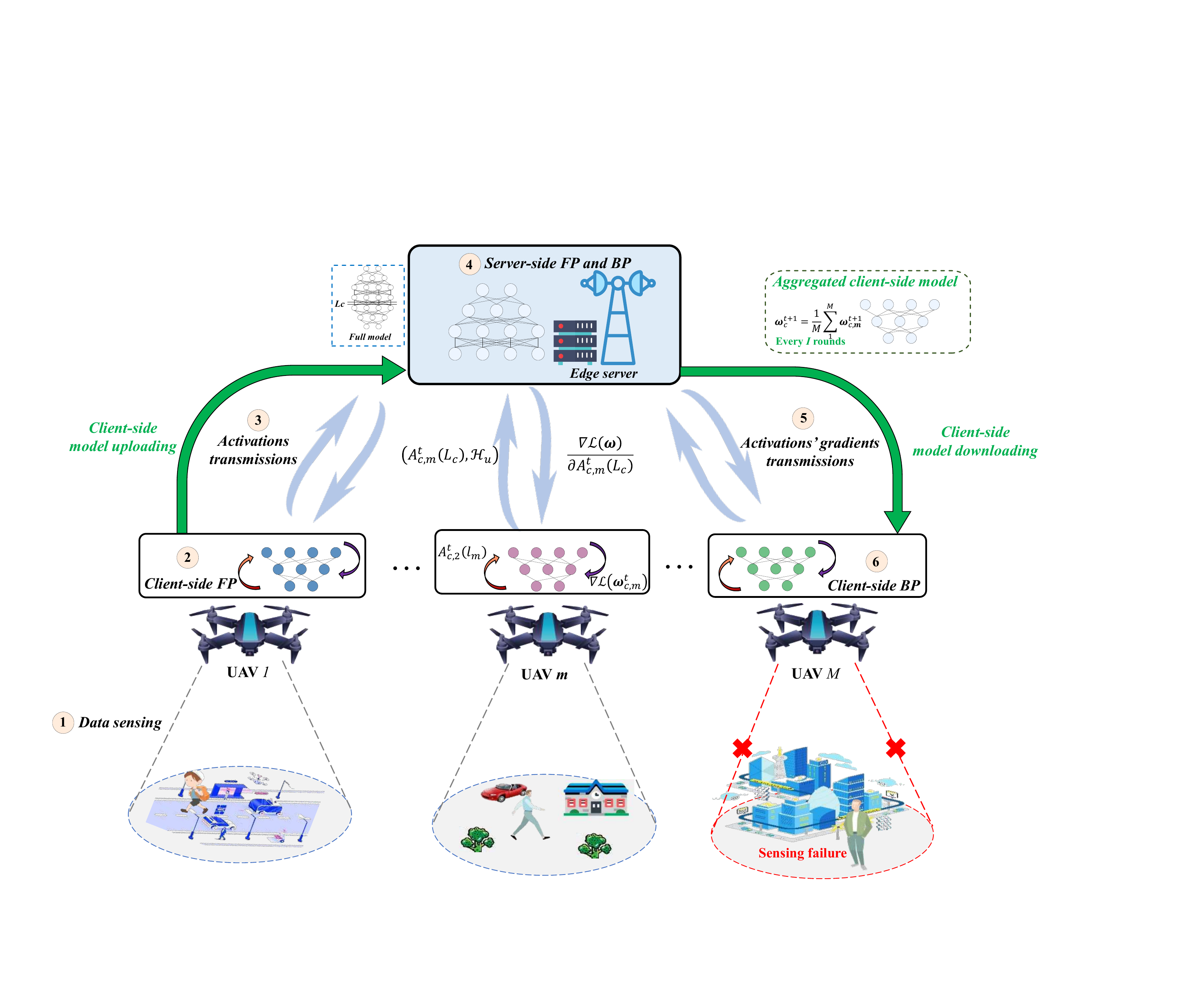} 
    \caption{The architecture of SFLSC3.}
    \label{FL1}
\end{figure}

The proposed SFLSC3 architecture is shown in Fig. \ref{FL1}, comprising an edge server, and a set of $M$ UAVs denoted as $\mathcal{M}=\{1,2, \ldots, m, \ldots, M\}$, is designed to collaboratively train a ML model for a specific sensing recognition task. Each UAV is equipped with a single-antenna ISAC transceiver that can switch between sensing and communication modes in a time-division manner depending on the mission phase. Specifically, during the sensing phase of round $t$, UAV $m$ senses the target by transmitting a frequency-modulated continuous wave (FMCW) and collect training data, which will be transmitted in the communication phase. However, considering the limits of sensing and communication resources, the full model $\boldsymbol{w}$ which has $L$-layers is divided into the client-side model $\boldsymbol{w}_{c,m}$ and the server-side model $\boldsymbol{w}_{s,m}$. The client-side model reserves neurons of layer $\{1,2,\dots,L_c\}$, while the server-side model reserves neurons of layer $\{L_c+1,L_c+2,\dots,L\}$, where $L_c$ represents the split layer. Meanwhile, let $\boldsymbol{u}_m= \left(x_m,y_m,H\right)$ and $\boldsymbol{u}_s= \left(x_s,y_s,z_s\right)$ denote the three-dimensional coordinates of UAV $m$ and the edge server, respectively.

\subsection{Split Federated Learning Model}
The training process is to implement a global loss minimization function through decentralized coordination of learning agents. Specifically, the global loss function to measure the training error can be denoted as
\begin{equation}
    \mathcal{L}\left(\boldsymbol{w}\right)=\frac{1}{M}\sum\limits_{m=1}^M\mathcal{L}_m\left(\boldsymbol{w}\right),
\end{equation}
where $\mathcal{L}_m\left(\boldsymbol{w}\right)\overset{\triangle}{=}\mathbb{E}_{\xi_m\sim\mathcal{D}_m}\left[l_m\left(\boldsymbol{w};\xi_m\right)\right]$ is the local loss function over the dataset $\mathcal{D}_m$, and $\xi_m$ is training randomness of local dataset $\mathcal{D}_m$ caused by the sampling of mini-batch data.  Thus, the whole training process of FL aims for solving the following optimization problem
\begin{equation}
	\boldsymbol{P} 1: \boldsymbol{w}^*=\operatorname{argmin} \mathcal{L}(\boldsymbol{w}) .
	\label{mecanum}
\end{equation}

For this purpose, the training process is performed iteratively in multiple rounds, and each round includes the following steps~\cite{B12}:

\subsubsection{Data sensing}
During the sensing phase of $t$-th round, UAV $m$ transmits specific FMCW to sense the target. For the target located at $\boldsymbol{v}_m=\left(x_{m,v},y_{m,v},0\right)$, the probability of successfully sensing by UAV $m$ can be denoted as \cite{ISAC-2306}
\begin{equation}
    q_{s,m}\left(\boldsymbol{u}_m\right)=\frac{1}{1+\varepsilon \exp \left(-\varsigma\left[\theta_{s, m}\left(\boldsymbol{u}_m\right)-\varepsilon\right]\right)},
\end{equation}
 where $\varepsilon$ and $\varsigma$ are constants corresponding to the environment \cite{6863654}, $\theta_{s, m}\left(\boldsymbol{u}_m\right)=\frac{180^{\circ}}{\pi} \times \arcsin |\frac{H}{d_{s,m}}|$ represents the elevation angle of target sensing, and $d_{s,m}=\|\boldsymbol{u}_m-\boldsymbol{v}_m\|$ denotes the distance between UAV $m$ and the server. 
 
 Upon successful sensing at $t$-th round, UAV $m$ acquires a batch of data samples of size $b$, represented as $\mathcal{D}_m^t=\left\{\left(\boldsymbol{x}_{m, 1}^t, \boldsymbol{y}_{m, 1}^t\right),\left(\boldsymbol{x}_{m, 2}^t, \boldsymbol{y}_{m, 2}^t\right), \ldots,\left(\boldsymbol{x}_{m, b}^t, \boldsymbol{y}_{m, b}^t\right)\right\}$. For the sample $\left(\boldsymbol{x}_{m, i}^t, \boldsymbol{y}_{m, i}^t\right)$, $\boldsymbol{x}_{m, i}^t \in \mathrm{R}^{N_{f \times 1}}$ and $\boldsymbol{y}_{m, i}^t \in \mathrm{R}^{N_{l \times 1}}$ represent the feature and corresponding label of the sample respectively. $N_f$ and $N_l$ represent the dimensions of the features and labels, respectively.

\subsubsection{Forward propagation of client-side}
At the $t$-th round, each UAV computes the activations to perform forward propagation (FP) of every layer $l_c\in\{1,\ldots,L_c\}$~\cite{B12}:
\begin{equation}    
A_{c,m}^{t}(l_c)=\delta\left(\boldsymbol{w}_{c,m}^{t}(l_c)A_{c,m}^{t}(l_c-1)+b_{c,m}^{t}(l_c)\right),
\end{equation}
where $\delta$ signifies the activation function, $\boldsymbol{w}_{c,m}^{t}(l_c)$ and $b_{c,m}^{t}(l_c)$ represent the weights and biases of the $l_c$-th layer, respectively.

The smashed data which involves the activation value $A_{c,m}^{t}(L_c)$ and its relevant label $\mathcal{H}_u$ can be denoted as
\begin{equation}
    \boldsymbol{S}_{c,m}^t=\left(A_{c,m}^{t}(L_c),\mathcal{H}_u\right),
\end{equation}

For the following-up training, the smashed data is uploaded to the edge server.

\subsubsection{Forward and backward propagation of server-side}
After gathering the smashed data $\boldsymbol{S}_{c,m}^t$ from each UAV, the edge server continues the forward propagation through the remaining layers $l_s\in \{L_c+1,L_c+2,\dots,L\}$
\begin{equation}
A_{s,m}^{t}(l_s)=\delta\left(\boldsymbol{w}_{s,m}^{t}(l_s)A_{s,m}^{t}(l_s-1)+b_{s,m}^{t}(l_s)\right),
\end{equation}
where $A_{s,m}^{t}(L_c)=A_{c,m}^{t}(L_c)$. Subsequently, the server determines the loss by comparing the final output to the labels provided by UAV $m$ and conducts backward propagation to calculate the gradients concerning the server's weights by
\begin{equation}
    \mathbf{g}\left(\boldsymbol{w}_{s,m}^t\right)=\nabla \mathcal{L}\left(\boldsymbol{w}_{s,m}^t\right)=\frac{\partial \mathcal{L}(\boldsymbol{w})}{\partial\boldsymbol{w}_{s,m}^t},
\end{equation}
and $\mathbf{g}\left(\boldsymbol{w}_{s,m}^t\right)=0$ if UAV $m$ fails to sense the target. Thus, the local model $\boldsymbol{w}_{s,m}^t$ is updated by
\begin{equation}
\boldsymbol{w}_{s,m}^{t+1}=\boldsymbol{w}_{s,m}^t-\eta\alpha_m^t{\mathbf{g}}\left(\boldsymbol{w}_{s,m}^t\right).
\end{equation}

The gradient of activation values are transmitted to each UAV and afterwards the edge server aggregates the server-side models at each round after updating them. Considering the server-side model for each UAV is all the same at each round, the server-side model is updated as
\begin{equation}
     \boldsymbol{w}_{s}^{t+1}=\frac{1}{M}\sum\limits_{m=1}^M\boldsymbol{w}_{s,m}^{t+1},
\end{equation}
where $\eta$ is a constant parameter denoting the learning rate.

\subsubsection{Client-side model updating}

After obtaining gradients $\mathbf{g}_{s,m}^t$ transmitted from the edge server, each UAV computes the gradients for client-side model through backward propagation by
\begin{equation}
    \mathbf{g}\left(\boldsymbol{w}_{c,m}^t\right)=\nabla \mathcal{L}\left(\boldsymbol{w}_{c,m}^t\right)=\frac{\partial \mathcal{L}(\boldsymbol{w})}{\partial A_{c,m}^t(L_c)}\cdot\frac{\partial A_{c,m}^{t}(L_c)}{\partial\boldsymbol{w}_{c,m}^t},
\end{equation}
while $\mathbf{g}\left(\boldsymbol{w}_{c,m}^t\right)=0$, if UAV $m$ fails to sense the target. Thus, the local model $\boldsymbol{w}_{c,m}^t$ performs update by
\begin{equation}
\boldsymbol{w}_{c,m}^{t+1}=\boldsymbol{w}_{c,m}^t-\eta\alpha_m^t{\mathbf{g}}\left(\boldsymbol{w}_{c,m}^t\right).
\label{modelcm}
\end{equation}

\subsubsection{Aggregation of client-side}
The UAVs send the client-side models $\boldsymbol{w}_{c,m}^{t+1}$ to the server every $I$ rounds for model aggregation, which can be denoted as
\begin{equation}
    \boldsymbol{w}_{c}^{t+1}=\frac{1}{M}\sum\limits_{m=1}^M\boldsymbol{w}_{c,m}^{t+1}.
    \label{modelc}
\end{equation}

Upon finishing the model aggregation, the server sends the model $\boldsymbol{w}_{c}^{t+1}$ which has been aggregated to all UAVs, and the model $\boldsymbol{w}_{c}^{t+1}$ will be used as the initial model for the next round.

\subsection{UAV Communication Model}
We model the communication between UAV $m$ and the server as a probabilistic combination of line-of-sight (LoS) and non-line-of-sight (NLoS) channels.The probability of LoS link can be given by
\begin{equation}
    q_{c,m}\left(\boldsymbol{u}_m\right)=\frac{1}{1+\varepsilon \exp \left(-\varsigma\left[\theta_{c, m}\left(\boldsymbol{u}_m\right)-\varepsilon\right]\right)},
\end{equation}
where $\theta_{c, m}\left(\boldsymbol{u}_m\right)=\frac{180^{\circ}}{\pi} \times \arcsin |\frac{z_m-z_s}{d_{c,m}}|$ represents the elevation angle of communication, and $d_{c,m}=\|\boldsymbol{u}_m-\boldsymbol{u}_s\|$ denotes the distance between UAV $m$ and the server. Thus, the probability of NLoS is $1-q_{c,m}\left(\boldsymbol{u}_m\right)$. Besides, the channel gain can be obtained by \cite{6863654}
\begin{equation}
    h_m\left(\boldsymbol{u}_m\right)=\frac{\left(\frac{4 \pi f_c}{c} d_{c, m}\right)^{-2}}{\gamma_{L o S} q_{c, m}\left(\boldsymbol{u}_m\right)+\gamma_{N L o S}\left(1-q_{c, m}\left(\boldsymbol{u}_m\right)\right)},
\end{equation}
where $\gamma_{L o S}$ and $\gamma_{N L o S}$ serve as the empirical path loss coefficients corresponding to LoS and NLoS links, respectively. Moreover, the data rate between UAV $m$ and the server is
\begin{equation}
R_m\left(\boldsymbol{u}_m\right)=B \log \left(1+\frac{p_c h_m\left(\boldsymbol{u}_m\right)}{B N_0}\right),
\end{equation}
where $B$ represents the communication bandwidth allocated for each UAV $m$, $p_c$ denotes the transmission power and $N_0$ signifies the noise power spectral density.

\subsection{Target Sensing Model}
The scattering from the target to UAVs can be simulated using the primitive-based method as described in~\cite{body}, where the target comprises $K$ body primitives. To enhance the signals scattered from the target and extract useful information, UAVs actively transmit FMCW signals consisting of multiple up-chirps~\cite{FMCW}. Let $s_m(t)$ represent the transmitted sensing signal of UAV $m$ at time $t$, the signal received by $m$-th UAV can be approximated as the superposition of returns scattered from $K$ body primitives, given by
\begin{equation} 
\begin{aligned}
y_m(t)&=\sqrt{p_s} \times \frac{A}{\sqrt{4 \pi}} \sum_{k=1}^K \frac{\sqrt{G_{m, k}(t)}}{d_{m, k}^2(t)} \times \\
&\exp \left(-j \frac{4 \pi f_c}{c} d_{m, k}(t)\right) s_{m, k}\left(t-2 \frac{d_{m, k}(t)}{c}\right)+n_m(t),
\end{aligned}
\end{equation}
where ${p_s}$ is UAV sensing transmit power, $A$ is the antenna gain, $G_{m, k}(t)$ denotes the complex amplitude that correlates with the radar cross section area of the $k$-th body primitive, $f_c$ is the carrier, $c$ is the speed of light, $d_{m, k}(t)$ is the distance between UAV $m$ and the $k$-th body primitive, and $n_m(t)$ is the interference signal caused by noise and environmental clutter.

The micro-Doppler characteristics produced by the different motion states of the target are different, and spectrogram~\cite{FMCW} is employed for micro-Doppler analysis in this paper. Each spectrogram is generated from the signal received within a time period $T_0=\lambda T_d$, which includes $\lambda$ chirps with a duration of $T_d$. Therefore, each UAV $m$ must persistently sense the target over a period $T_{m,s}=b T_0$ to acquire $b$ data samples (i.e., spectrograms).

\subsection{Delay Model}

At the $t$-th round, the delay of UAV $m$ can be given as follows~\cite{B7}.

\subsubsection{Delay of target sensing}
The delay consumed by UAV $m$ for target sensing to collect data can be denoted as $T_{m,s}$.

\subsubsection{Delay of local training}
Let $\xi_f$ and $\xi_b$ be the computation workload, evaluated in floating point operations per second (FLOPS), of forward propagation and backward propagation for a single data sample, respectively. The local training latency for the client-side model $\boldsymbol{w}_{c,m}^t$ can be given by
\begin{equation}
    T_{m,tr}=\frac{b(\xi_{f,L_c}+\xi_{b,L_c})}{f_m\varpi}
\end{equation}
where $f_m$ signifies the computing capability (i.e., CPU frequency) of UAV $m$, and $\varpi$ denotes the computing intensity. 

\subsubsection{Delay of communication}
The transmission latency to upload the smashed data $\boldsymbol{S}_{c,m}$ of UAV $m$ can be given as
\begin{equation}
    T_{m,com}^{sma}=\frac{\boldsymbol{S}_{c,m}}{R_m\left(\boldsymbol{u}_m\right)},
\end{equation}

Meanwhile, the transmission latency to upload local model parameters $|\boldsymbol{w}_{c,m}|$ can be represented by
\begin{equation}
    T_{m,com}^{par}=\frac{\boldsymbol{d}_m\left(L_c\right)}{R_m\left(\boldsymbol{u}_m\right)}.
\end{equation}
And $\boldsymbol{d}_m\left(L_c\right) = \boldsymbol{A}\left(\frac{L_c}{L}\right)^2m_{\boldsymbol{w}}$, where $\boldsymbol{A}$ is a positive constant and $m_{\boldsymbol{w}}$ denotes the bits of overall model parameters.
\subsubsection{Overall delay of $I$ rounds}
We consider each round has $e$ epochs, thereby the delay of UAV $m$  can be given by
\begin{equation}
    T_{m,t}^{epo}=\alpha_m^t e\left(T_{m,tr}+T_{m,com}^{sma}+T_{m,s}\right).
\end{equation}

Besides, given that the split training is carried out every round while the local model parameters uploading occurs every $I$ rounds, thus the overall expected delay of UAV $m$ for $I$ rounds can be calculated by
\begin{equation}
    T_m\left(I,L_c,\boldsymbol{u}_m,b\right)=\mathbb{E}[IT_{m,t}^{epo}+T_{m,com}^{par}],
\end{equation}
where $\mathbb{E}[\cdot]$ denotes the mathematical expectation about $t$.
\section{Convergence Analysis}
In this section, we evaluate the impact of split point selection, aggregation frequency, data sensing volume and locations of UAVs on the proposed SFLSC3 through conducting a detailed convergence analysis.

\subsection{Basic Assumptions}
Prior to embarking on the convergence analysis, several widely accepted assumptions are presented as follows:

\textbf{Assumption 1.} \textit{Each local loss function $\mathcal{L}_m\left(\boldsymbol{w}\right)$ is uniformly $\beta$-smooth with respect to any $\boldsymbol{w}$ and $\boldsymbol{w}^{\prime}$, thus we have}
\begin{equation}
    \left\|\nabla \mathcal{L}
    _m\left(\boldsymbol{w}\right)-\nabla \mathcal{L}_m\left(\boldsymbol{w}^{\prime}\right)\right\|\leq \beta\left\|\boldsymbol{w}-\boldsymbol{w}^{\prime}\right\|, \forall m,
\end{equation}
\textit{where $\beta$ represents the Lipschitz constant corresponding to $\mathcal{L}\left(\cdot\right)$}.

\textbf{Assumption 2.} \textit{The variance and second moments of stochastic gradients for each layer are bounded by}
\begin{equation}
\mathbb{E}\left[\Vert \mathbf{g}_m - \nabla \mathcal{L}_m(\boldsymbol{w}) \Vert^2 \right] \leq \frac{\sum\limits_{l=1}^L \sigma_l^2}{b}, \forall m, \forall \boldsymbol{w},
\end{equation}
\textit{and}
\begin{equation}
\mathbb{E}\left[\Vert \mathbf{g}_m \Vert^2 \right] \leq\sum\limits_{l=1}^{L_c}G_l^2, \forall m, \forall \boldsymbol{w},
\end{equation}
\textit{where $\sigma_l^2$ and $G_l^2$ denote the variance and second moments' upper bound for l-th layer of model $\boldsymbol{w}$, respectively.}

\textbf{Assumption 3.} \textit{The heterogeneity which exists in the local datasets can by assessed by}
\begin{equation}
    \mathbb{E}\left[\left\|\nabla\mathcal{L}_m\left(\boldsymbol{w}\right)-\nabla\mathcal{L}\left(\boldsymbol{w}\right)\right\|^2\right]\leq \Lambda_m^2
\end{equation}
\textit{where $\Lambda_l^2$ represents a positive constant.}

\textbf{Assumption 4.} \textit{The stochastic gradients exhibit unbiasedness, which can be denoted as}
\begin{equation}
    \mathbb{E}\left[\mathbf{g}_m\right]=\nabla\mathcal{L}_m\left(\boldsymbol{w}\right).
\end{equation}

\subsection{Convergence Analysis}
Before establishing the convergence rate, some key Lemmas are presented as follows.
\begin{Lemma}
According to \textbf{Assumption 1}, we can derive
\begin{equation}
    \mathbb{E}\left[\left\|\boldsymbol{w}_c^{t}-\boldsymbol{w}_{c,m}^{t}\right\|^2\right] \leq 4\eta^2I^2\sum\limits_{l=1}^{L_c}G_l^2.
\end{equation}
\begin{proof}
See APPENDIX A.
\end{proof}
\end{Lemma}

\begin{Lemma}
Under \textbf{Assumption 1} and \textbf{Lemma 1}, we have
\begin{equation}
\begin{aligned}
&\mathbb{E}\left[\langle \nabla \mathcal{L}\left(\boldsymbol{w}^{t-1}\right),\boldsymbol{w}^t-\boldsymbol{w}^{t-1} \rangle\right]\\
&\leq
\eta \sum\limits_{m=1}^M \varphi_m^2 \left(\sum\limits_{m=1}^M \Lambda_m^2 +4M\beta^2\eta^2I^2\sum\limits_{l=1}^{L_c}G_l^2\right)\\
&- \frac{\eta}{2} \mathbb{E}\left[\left\|\sum\limits_{m=1}^M \varphi_m \nabla \mathcal{L}\left(\boldsymbol{w}^{t-1}\right)\right\|^2\right].
\end{aligned}
\end{equation}

\begin{proof}
See APPENDIX B.
\end{proof}
\end{Lemma}

\begin{Lemma}
According to \textbf{Assumption 3} and \textbf{Assumption 4}, it holds that
\begin{equation}
\begin{aligned}
&\mathbb{E}\left[\|\boldsymbol{w}^t-\boldsymbol{w}^{t-1}\|^2\right] < \frac{2\eta^2}{M}\sum\limits_{m=1}^M \varphi_m\frac{\sum\limits_{l=1}^{L_c} \sigma_l^2}{b}+\frac{4\eta^2}{M^2} \sum\limits_{m=1}^M \Lambda_m^2\\&+4\eta^2\sum_{m=1}^M\kappa_m\left((q_{s,m}-\bar{q}_s)^2+\bar{q}_s^2\right)\Lambda_m^2\\&+32\beta^2\eta^4I^2\sum\limits_{l=1}^{L_c}G_l^2+8\eta^2\mathbb{E}\left[\|\nabla\mathcal{L}(\boldsymbol{w}^{t-1})\|^2\right].
\end{aligned}
\end{equation}
\begin{proof}
 See APPENDIX C.   
\end{proof}
\end{Lemma}

\begin{theorem}
We consider the learning rate $\eta$ of proposed SFLSC3 satisfies that
\begin{equation}
0<\eta<\frac{\sum\limits_{m=1}^M\varphi_m}{8\beta}.
\label{etacon}
\end{equation}

Under \textbf{Assumption} 1-4 and \textbf{Lemma} 1-3, we can obtain
\begin{equation}
\begin{aligned}
&\frac{1}{N}\sum\limits_{t=1}^N\mathbb{E}\left[\|\nabla\mathcal{L}(\boldsymbol{w}^{t-1})\|^2\right]\leq\frac{2}{\eta\left(\sum\limits_{m=1}^M\varphi_m\right)^2-8\beta\eta^2}\cdot\\&\quad\Bigg(\frac{\mathbb{E}\left[\mathcal{L}\left(\boldsymbol{w}^{0}\right)\right]-\mathbb{E}\left[\mathcal{L}\left(\boldsymbol{w}^{*}\right)\right]}{N}+\frac{\sum\limits_{l=1}^{L} \beta\eta^2\sigma_l^2}{Mb}\\& \quad+\frac{2\beta\eta^2}{M^2} \sum\limits_{m=1}^M \Lambda_m^2+16\beta^3\eta^4I^2\sum\limits_{l=1}^{L_c}G_l^2\\&\quad+2\beta\eta^2\sum_{m=1}^M\kappa_m\left((q_{s,m}-\bar{q}_s)^2+\bar{q}_s^2\right)\Lambda_m^2\\&\quad+  \sum\limits_{m=1}^M \eta\Lambda_m^2 +4M\beta^2\eta^3I^2\sum\limits_{l=1}^{L_c}G_l^2\Bigg).
\end{aligned}
\label{theorem}
\end{equation}

\begin{proof}
    See APPENDIX D.
\end{proof}

Eq. \eqref{theorem} indicates that when the UAVs have varying target sensing probabilities, the adverse effects resulting from data heterogeneity $2\beta\eta^2\sum_{m=1}^M\kappa_m\left((q_{s,m}-q_s)^2+q_s^2\right)\Lambda_m^2$ will be amplified. For alleviating this adverse effects, we assume the UAVs have the uniform target sensing probabilities, thus $q_{s,m}=q_s$. Moreover, we can derive the following corollary.
\end{theorem}

\begin{corollary}
Under the constraint of learning rate $\eta$ as shown in Eq. \eqref{etacon}, when the UAVs have uniform target sensing probabilities, we can obtain the expression of $\varphi_m$ and $\kappa_m$, which can be given by
\begin{equation}
\begin{aligned}
\varphi_m = \frac{q_s}{\left[1-(1-q_s)^M\right]M}.
\end{aligned}
\end{equation}
\begin{proof}
    See APPENDIX E.
\end{proof}
\begin{equation}
    \kappa_m = \frac{2}{Mq_s^M}.
\end{equation}
\begin{proof}
    See APPENDIX B of \cite{ISAC-2306}.
\end{proof} 
\end{corollary}
\begin{corollary}
When $\frac{1}{N}\sum\limits_{t=1}^{N} \mathbb{E}\left[\|\nabla \mathcal{L}\left(\mathbf{w}^{t-1}\right)\|^{2}\right] \leq \varepsilon$, it represents the SFLSC3 has converged to the target convergence accuracy $\varepsilon$, and the training rounds demanded can be represented as
\begin{equation}
\begin{aligned}
N \geq \frac{\vartheta}{\varepsilon^{\prime}-\Gamma_1\sum\limits_{l=1}^{L_c}G_l^2I^2-\frac{\Gamma_2}{b}-\frac{\Gamma_3}{q_s^{M-2}}-\Gamma_4}
\end{aligned}
\end{equation}
with
\begin{subequations}
\begin{align}
&\vartheta=\mathbb{E}\left[\mathcal{L}\left(\boldsymbol{w}^{0}\right)\right]-\mathbb{E}\left[\mathcal{L}\left(\boldsymbol{w}^{*}\right)\right],\\
&\varepsilon^{\prime}=\eta\varepsilon\left(\frac{q_s}{4\left[1-(1-q_s)^M\right]M}\right)^2-4\varepsilon\beta\eta^2,\\
&\Gamma_1=16\beta^3\eta^4+4M\beta^2\eta^3,\\
&\Gamma_2=\frac{\sum\limits_{l=1}^{L} \beta\eta^2\sigma_l^2}{M},\\
&\Gamma_3=8U\beta^2\gamma^2,\\
&\Gamma_4=\sum\limits_{m=1}^M \eta\Lambda_m^2+\frac{2\beta\eta^2}{M^2} \sum\limits_{m=1}^M \Lambda_m^2.
\end{align}
\end{subequations}
\end{corollary}

\section{Problem Formulation and Solution
Algorithm}
As previously highlighted, balancing training performance and delay is critical due to the limited battery capacity of UAVs. To address this, we formulate an delay minimization problem with convergence guarantees by jointly optimizing the aggregation frequency \( I \), split point selection \( L_c \), UAV 3-D coordinates \( \{\boldsymbol{u}_m\} \), and data sensing volume \( b \).
\begin{subequations}
\begin{align}
\mathcal{P} 1: & \min_{I,L_c,\{\boldsymbol{u}_m\},b} \sum_{m=1}^M \mathcal{E} T_m(I,L_c,\boldsymbol{u}_m,b) \\
 \text{s.t.} & \quad \frac{1}{N} \sum_{n=1}^{N} \mathbb{E}[\|\nabla \mathcal{L}(\boldsymbol{w}_m)\|^2] \leq \varepsilon, \label{eq35b}\\
& \quad 0< L_c \leq L, \label{eq35c}\\
& \quad q_{s,m}\left(\boldsymbol{u}_m\right)=q_{s,m^{\prime}}\left(\boldsymbol{u}_{m^{\prime}}\right), \forall m,m^{\prime} \in \mathcal{M},\label{qscon}\\
& \quad \theta_{s}\left(\boldsymbol{u}_m\right)\geq \theta_0, \forall m \in \mathcal{M},\label{thetascon}\\
& \quad I \in \mathbb{N}^{+},\label{eq35e}\\
& \quad b \in \mathbb{N}^{+}\label{bfan},
\end{align}
\end{subequations}
where \( \mathcal{E} = \frac{N}{I} \) represents the total number of communication cycles required to achieve model convergence. Eq. \eqref{eq35b} ensures that the number of training rounds \( N \) satisfies the global convergence accuracy, while Eq. \eqref{eq35c} specifies the valid range of the split point \( L_c \). The constraint Eq. \eqref{qscon} originates from \textbf{Corollary 1}, which requires that UAVs maintain a consistent target sensing probability. To ensure reliable target sensing quality, we impose the constraint in Eq. \eqref{thetascon}. Eqs. \eqref{eq35e} and \eqref{bfan} ensure that the aggregation frequency \( I \) and data sensing volume \( b \) are positive integers.

Based on \textbf{Corollary 2}, we set $N$ to its minimum value and explicitly express $T$. Thus, problem $\mathcal{P}1$ can be rewritten as
\begin{subequations}
\begin{align}
\mathcal{P} 2: & \min_{I,L_c,\{\boldsymbol{u}_m\},b} \Xi\left(I,L_c,\{\boldsymbol{u}_m\},b\right)\\
 \text{s.t.} & \quad \rm Eq.~\eqref{eq35c}- \rm Eq.~\eqref{bfan}\label{eq37b},
\end{align}
\end{subequations}
where
\begin{equation}
    \Xi\left(I,L_c,\{\boldsymbol{u}_m\},b\right)=\frac{\sum\limits_{m=1}^M\vartheta E_m(I,L_c,\boldsymbol{u}_m,b)}{I\left(\varepsilon^{\prime}-\Gamma_1\sum\limits_{l=1}^{L_c}G_l^2I^2-\frac{\Gamma_2}{b}-\frac{\Gamma_3}{q_s^{M-2}}-\Gamma_4\right)}.
\label{Xi}
\end{equation}

Evidently, problem $\mathcal{P} 2$ exhibits non-convex characteristics due to the presence of both coupled integer and continuous variables. To address this, we decompose the problem into four sub-problems. When the split point $L_c$, 3-D coordinates of UAVs $\{\boldsymbol{u}_m\}$, and data sensing volume $b$ are fixed, problem $\mathcal{P} 2$ reduces to
\begin{subequations}
\begin{align}
\mathcal{P} 3: & \min_{I} \Xi\left(I\right) \\
 \text{s.t.} &   \quad \rm Eq. ~ \eqref{eq35e},
\end{align}
\end{subequations}

Then we can easily calculate the optimal client-side aggregation frequency $I^*$ by Newton Raphson's method, which is
\begin{equation}\label{I*}
I^*=\left\{\begin{array}{cc}1&I^{\prime}\leq1\\\arg\min_{I\in\{\lfloor I^{\prime}\rfloor,\lceil I^{\prime}\rceil\}}\Xi\left(I\right)&I^{\prime}>1\end{array}\right.
\end{equation}
where $I^{\prime}$ is an an extreme point of the derivative of $\Xi\left(I\right)$.

Similarly, when aggregation frequency $I$, split point $L_c$ and 3-D coordinates of UAVs $\{\boldsymbol{u}_m\}$ are fixed, the problem $\mathcal{P} 2 $ becomes
\begin{subequations}
\begin{align}
\mathcal{P} 4: & \min_{b} \Xi\left(b\right) \\
 \text{s.t.} &  \quad \rm Eq.~ \eqref{bfan}.
\end{align}
\end{subequations}

Next, the optimal data sensing volume $b$ can be given by
\begin{equation}
b^*=\left\{\begin{array}{cc}1&b^{\prime}\leq1\\\arg\min_{b\in\{\lfloor b^{\prime}\rfloor,\lceil b^{\prime}\rceil\}}\Xi\left(b\right)&b^{\prime}>1\end{array},\right.
\end{equation}
where $b^{\prime}$ is an an extreme point of the derivative of $\Xi\left(b\right)$.

Moreover, when we fix the aggregation frequency $I$, UAV 3-D coordinates $\{\boldsymbol{u}_m\}$ and data sensing volume $b$, the problem $\mathcal{P} 2$ can be rewritten as
\begin{subequations}
\begin{align}
\mathcal{P} 5: & \min_{L_c} \Xi\left(I^{*},L_c,\{\boldsymbol{u}_m\},b^{*}\right)\\
 \text{s.t.} &  \quad \rm Eq.~\eqref{eq35c}.
\end{align}
\end{subequations}

Given that the split point $L_c$ is discrete and its variable space is small, the optimal split point $L_c^{*}$ can be directly obtained using the traversing method. Specifically, the split point $L_c^{*}$ that yields the minimum value of the problem $\mathcal{P} 5$ is selected as the optimal solution.
Finally, based on $I^*$, $L_c^*$ and $b^*$, the problem $\mathcal{P} 2$ can be converted into

\begin{subequations}
\begin{align}
\mathcal{P} 6: & \min_{\{\boldsymbol{u}_m\}} \Xi\left(I^{*},L_c^*,\{\boldsymbol{u}_m\},b^{*}\right)\\
 \text{s.t.} &  \quad \rm Eq.~\eqref{thetascon}.
\end{align}
\end{subequations}

According to \textbf{\textit{Proposition 1}} in \cite{ISAC-2306}, when given $q_s$, the optimal position $\boldsymbol{u}_m^{*}$ of each UAV $m$ to maximize $R_m\left(\boldsymbol{u}_m\right)$ and then minimize $\Xi\left(I^{*},L_c^*,\boldsymbol{u}_m,b^{*}\right)$ can be derived from
\begin{equation}
    x_{m}^{*}=x_{m,v}-\frac{H(x_{m,v}-x_{s})}{\tan\theta_{s}\|\boldsymbol{u}_s-\boldsymbol{u}_m\|}
\end{equation}
and
\begin{equation}
    y_{m}^{*}=y_{m,v}-\frac{H(y_{m,v}-y_{s})}{\tan\theta_{s}\|\boldsymbol{u}_s-\boldsymbol{u}_m\|}.
\end{equation}

Based on $\boldsymbol{u}_m$ with given $q_s$, the transmission rate $R_m\left(\boldsymbol{u}_m\right)$ can be converted into $\hat{R}_m\left(q_s\right)$, and the problem $\mathcal{P} 6$ can be rewritten as
\begin{subequations}
\begin{align}
\mathcal{P} 6^{\prime}: & \min_{q_s} \Xi\left(I^{*},L_c^*,q_s,b^{*}\right)\\
 \text{s.t.} &  \quad \rm Eq.~\eqref{thetascon}.
\end{align}
\end{subequations}
which can be solved by SCA algorithm and get optimal $q_s^*$.  And based on this, we
employ the iterative block-coordinate descent(BCD)-based
algorithm to solve problem $\mathcal{P} 2$.
    

\begin{figure}[t]
    \centering
    \includegraphics[width=0.4\textwidth]{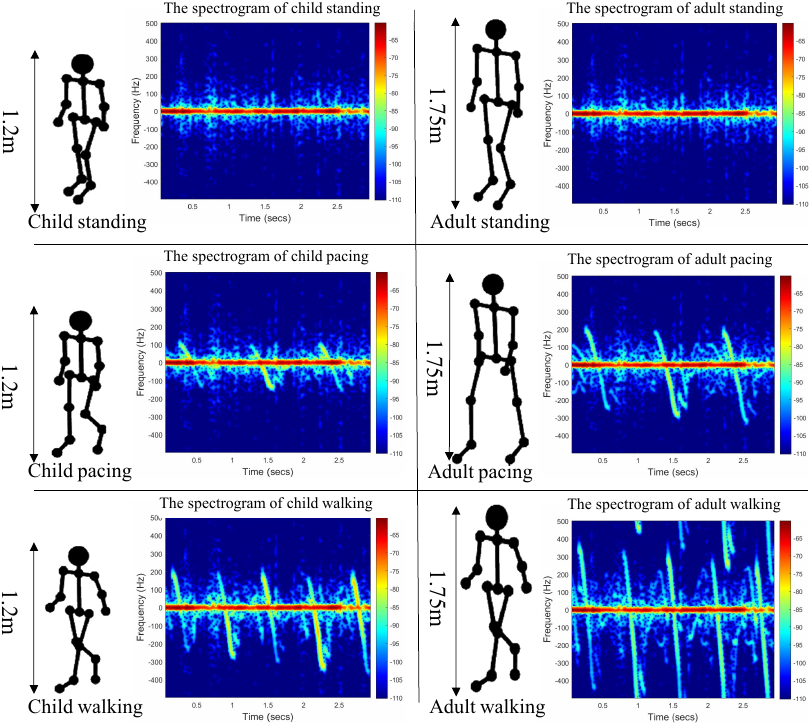} 
    \caption{Spectrograms of different motions generated from simulator\cite{9593198}.}
    \label{dataset}
\end{figure}
\begin{figure*}[!t]
	\centering  
	\subfigbottomskip=1pt 
	\subfigcapskip=-5pt 
	\subfigure[Total Delay.]{
		\includegraphics[width=0.32\linewidth]{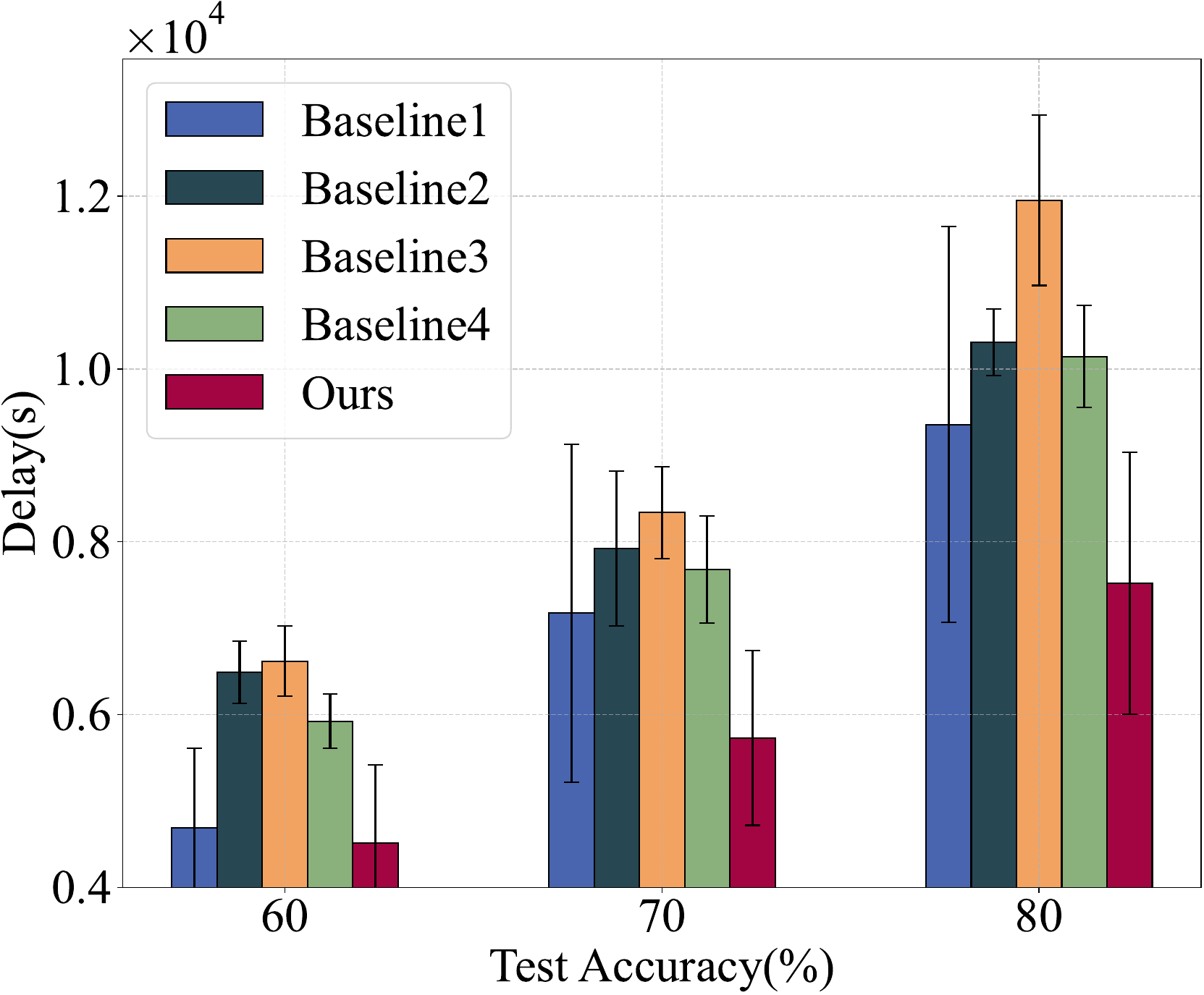}\label{3a}}
	\subfigure[Average test accuracy vs round.]{
		\includegraphics[width=0.32\linewidth]{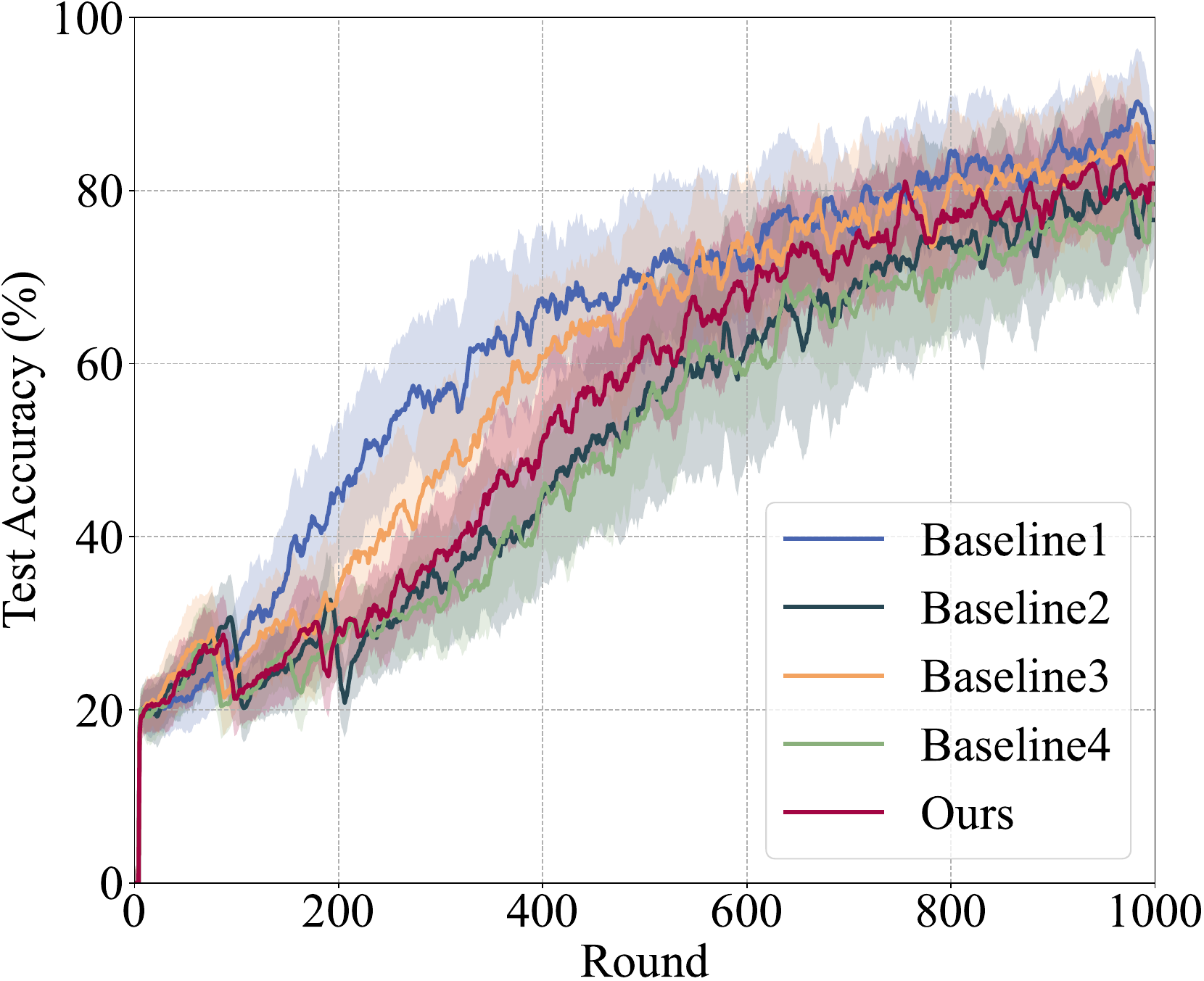}\label{3b}}
        \subfigure[Average test accuracy vs time.]{
		\includegraphics[width=0.32\linewidth]{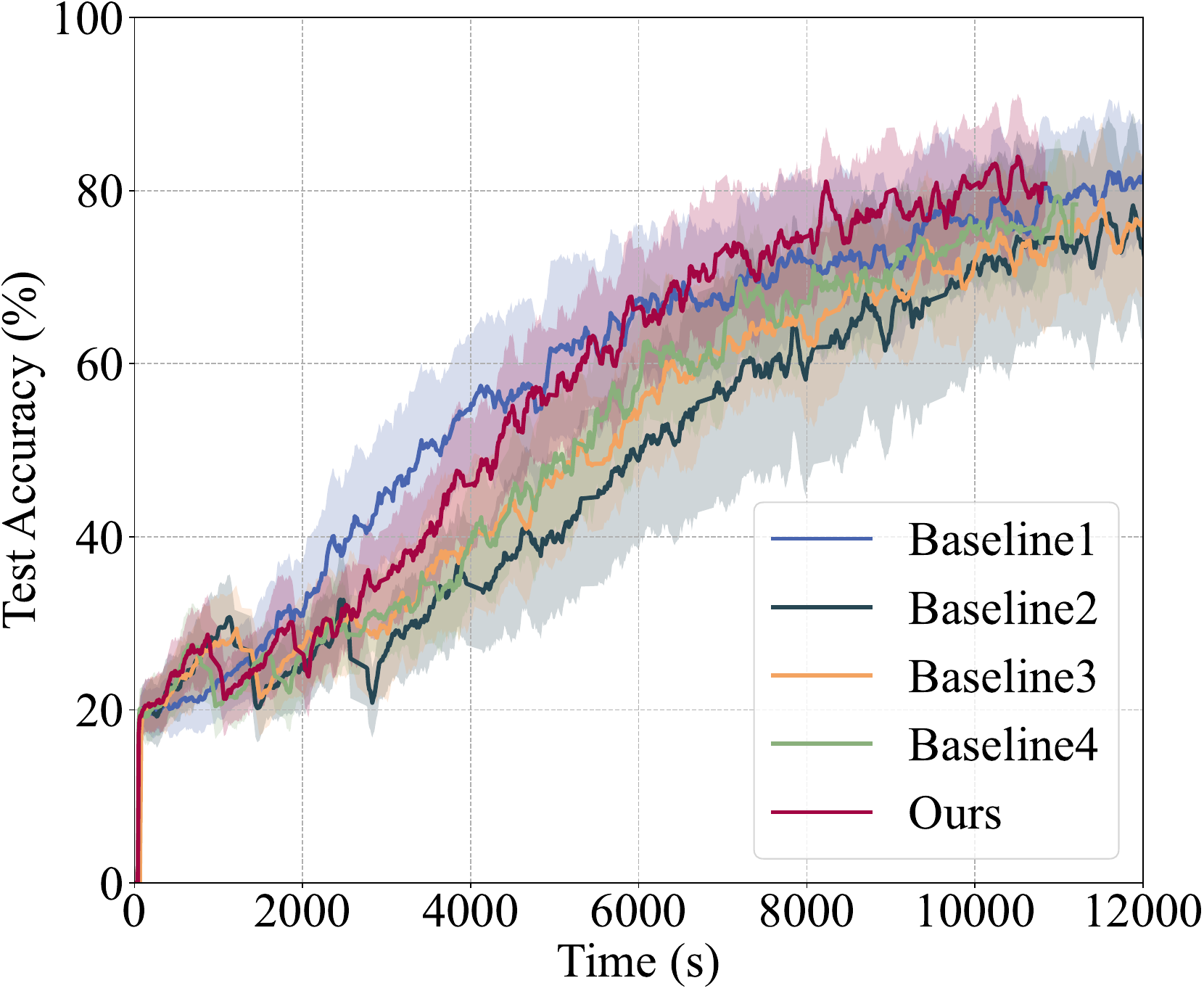}\label{3c}}
	\caption{Comparison of delay and convergence performance across different schemes.}
 \label{fig.3}
\end{figure*}
\begin{figure*}[t]
	\centering  
	\subfigbottomskip=1pt 
	\subfigcapskip=-5pt 
	\subfigure[SFLSC3 with different $I$.]{
		\includegraphics[width=0.23\linewidth]{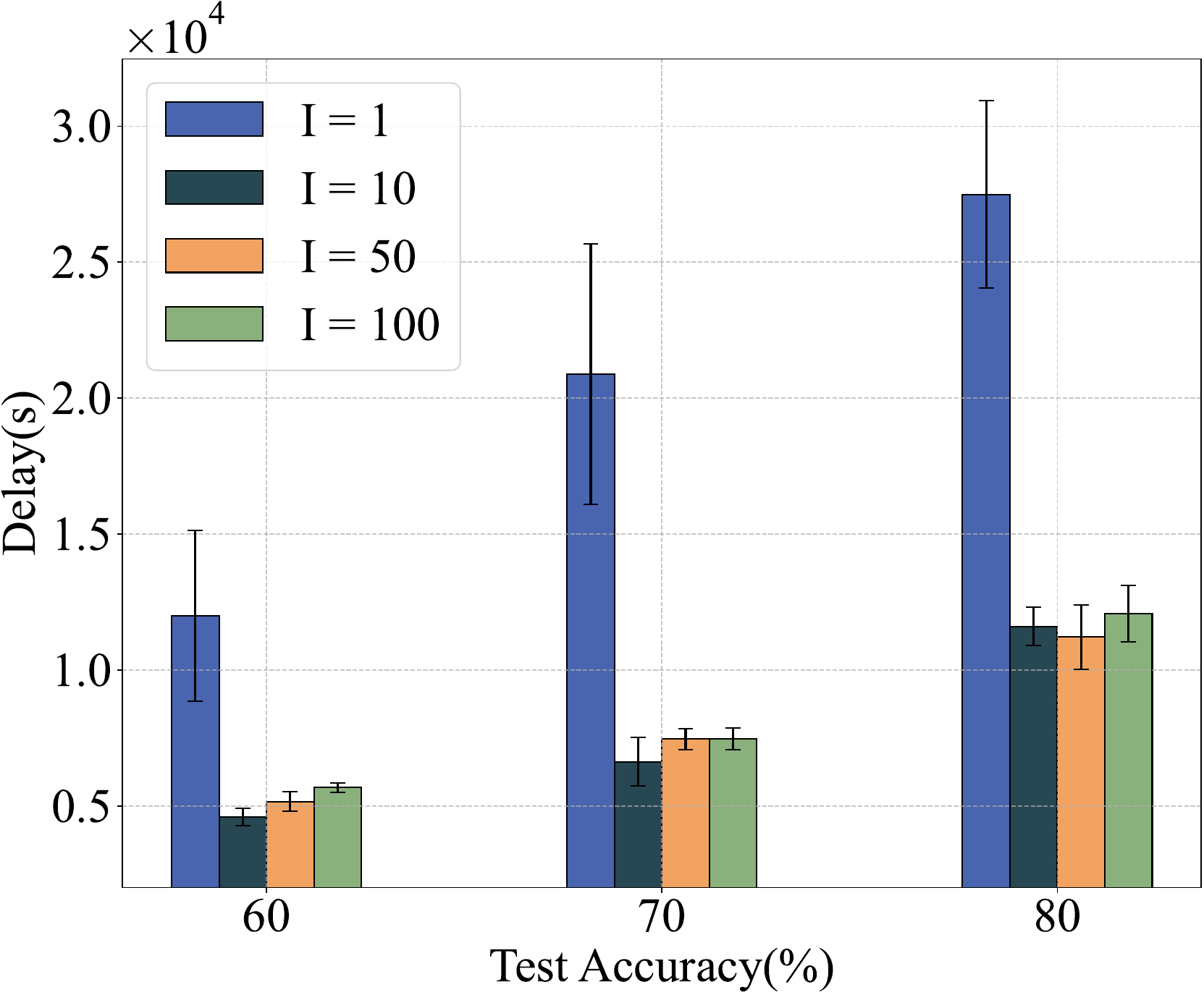}\label{4a}}
	\hspace{2pt} 
	\subfigure[SFLSC3 with different $L_c$.]{
		\includegraphics[width=0.23\linewidth]{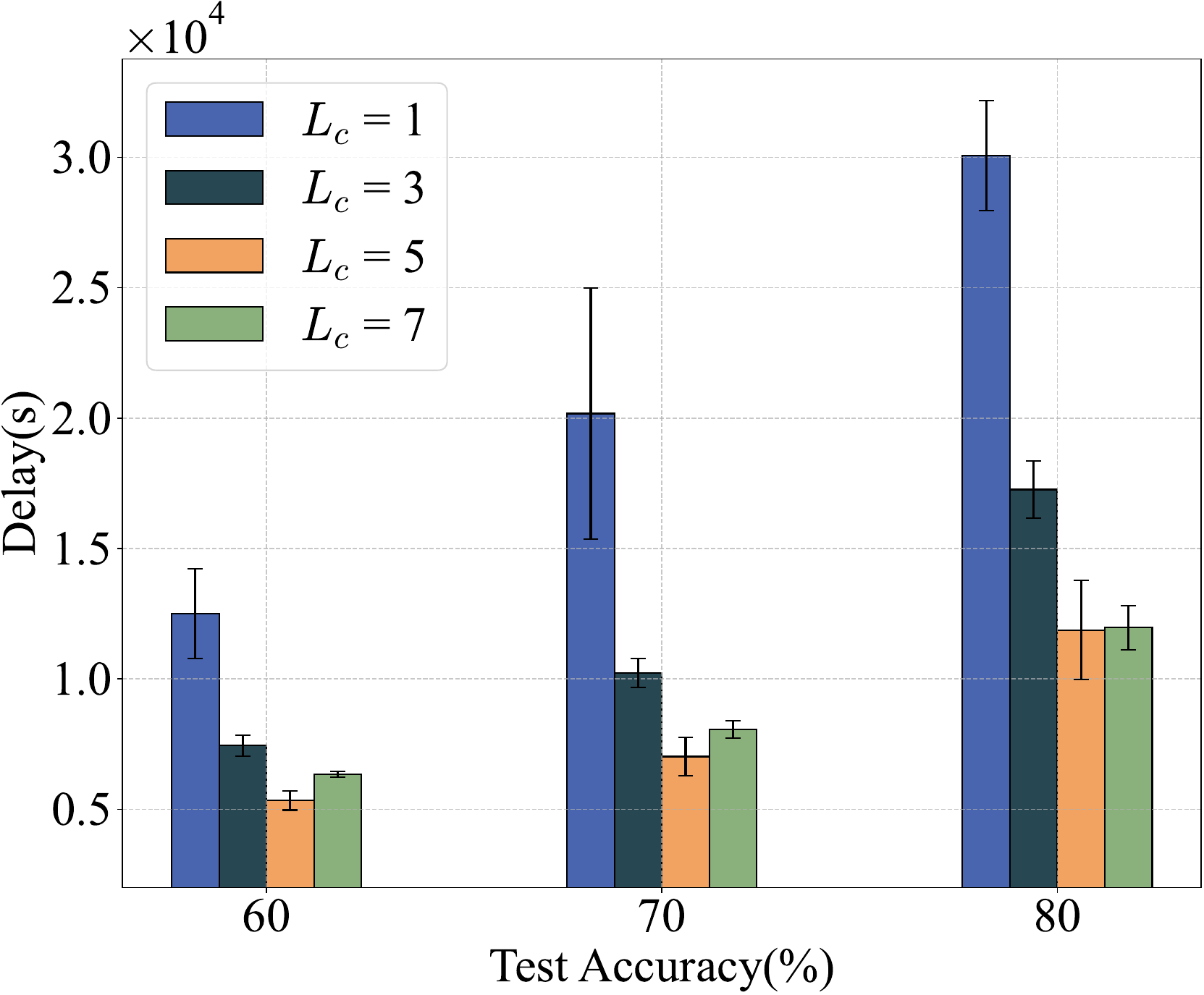}\label{4b}}
	\hspace{2pt}
    \subfigure[SFLSC3 with different $b$.]{
		\includegraphics[width=0.23\linewidth]{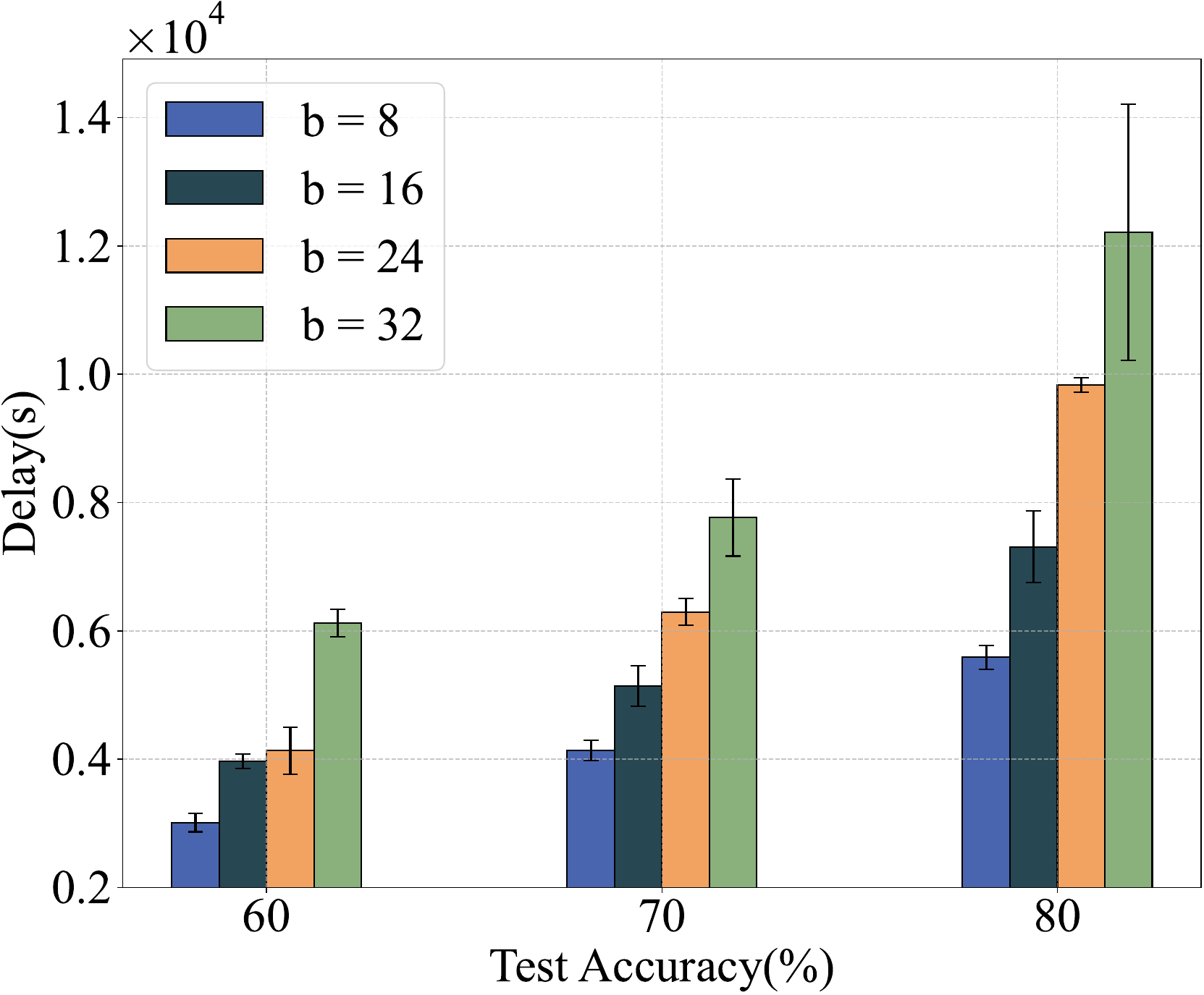}\label{4c}}
	\hspace{2pt}
    \subfigure[SFLSC3 with different $\theta_s$.]{ 
		\includegraphics[width=0.23\linewidth]{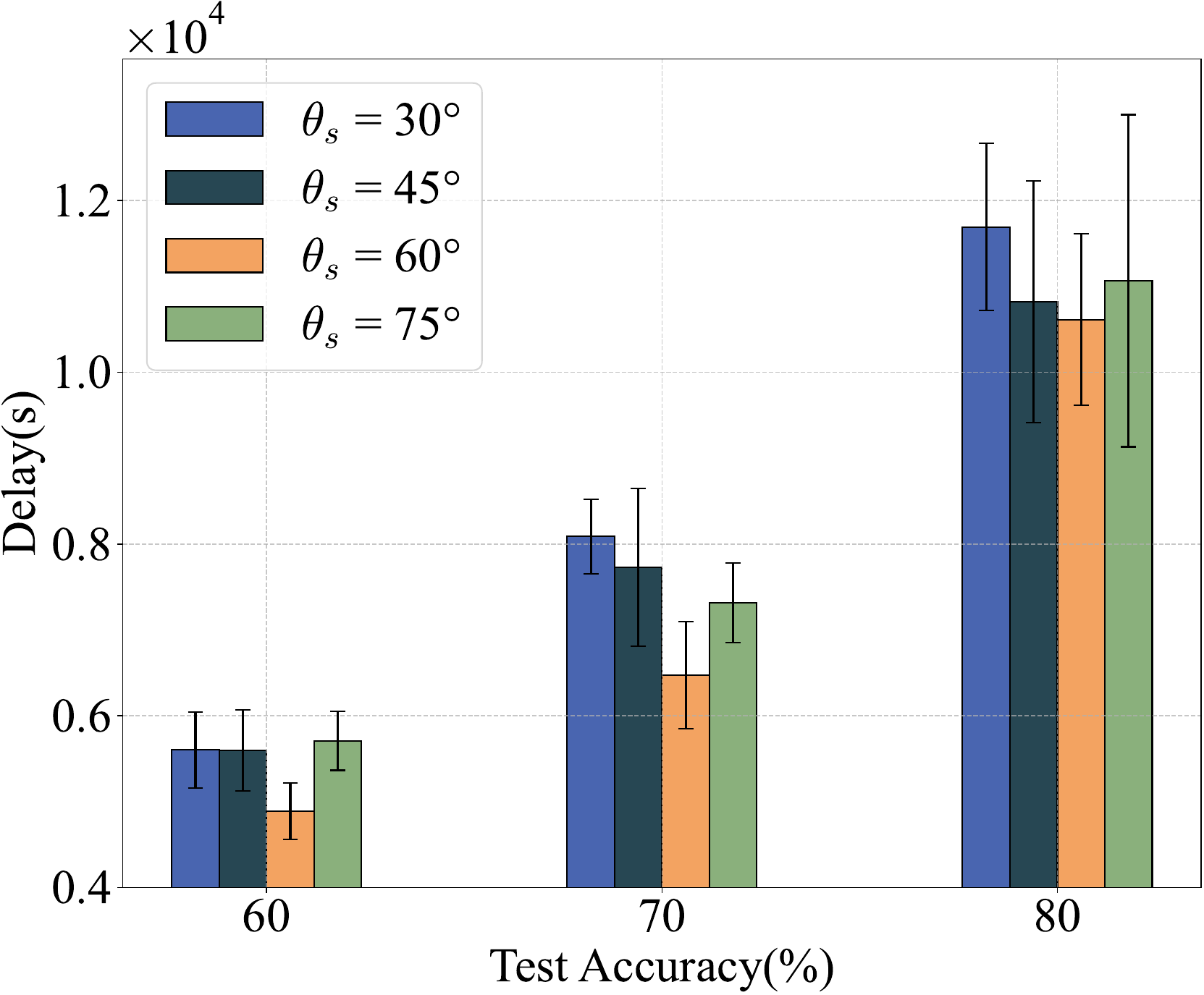}\label{4d}}
	\caption{The impact of different optimization variables on delay of SFLSC3.}
 \label{fig.4}
\end{figure*}

\begin{figure*}[t]
	\centering  
	\subfigbottomskip=1pt 
	\subfigcapskip=-5pt 
	\subfigure[SFLSC3 with different $I$.]{
		\includegraphics[width=0.23\linewidth]{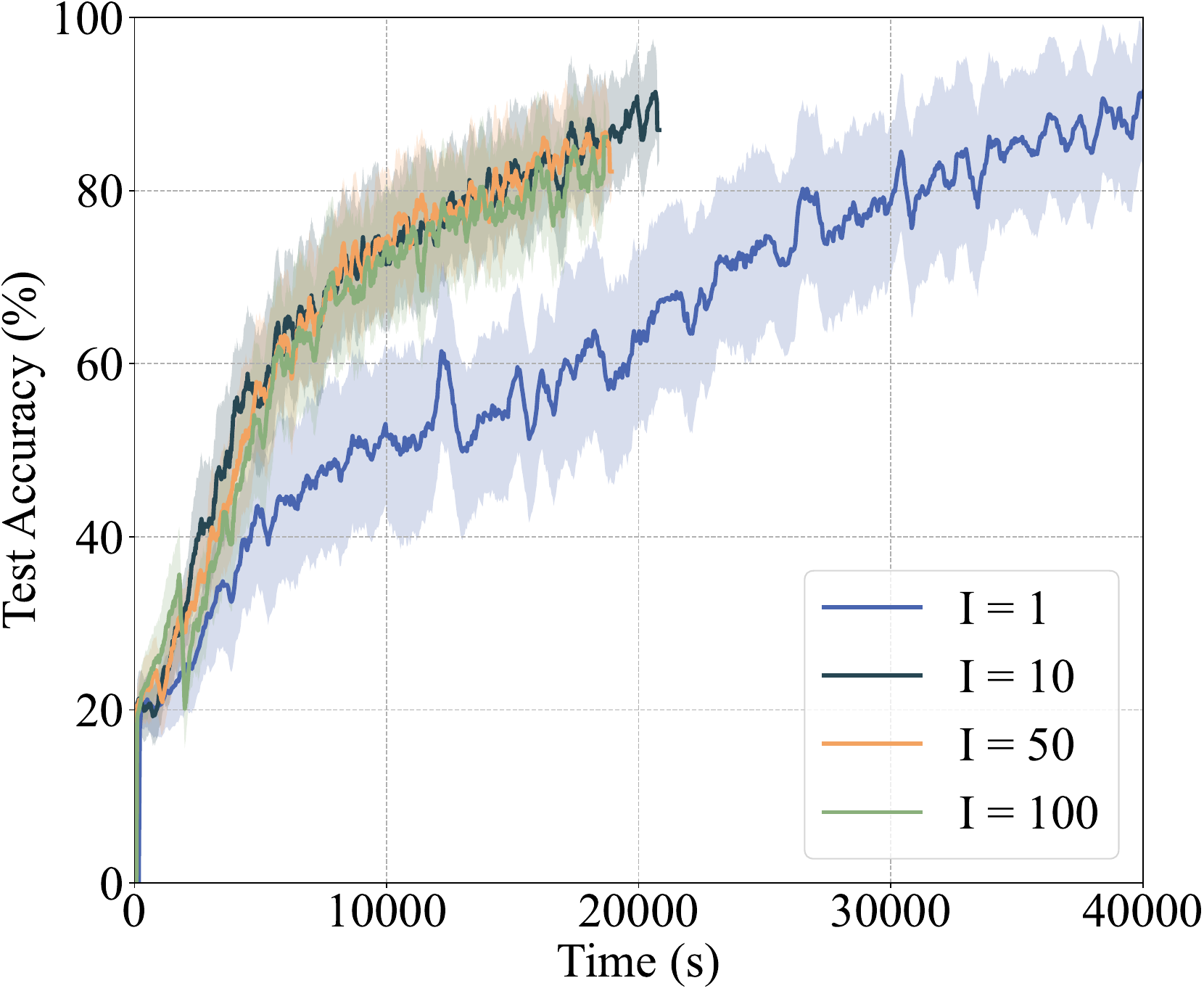}\label{5a}}
	\hspace{2pt}
	\subfigure[SFLSC3 with different $L_c$.]{
		\includegraphics[width=0.23\linewidth]{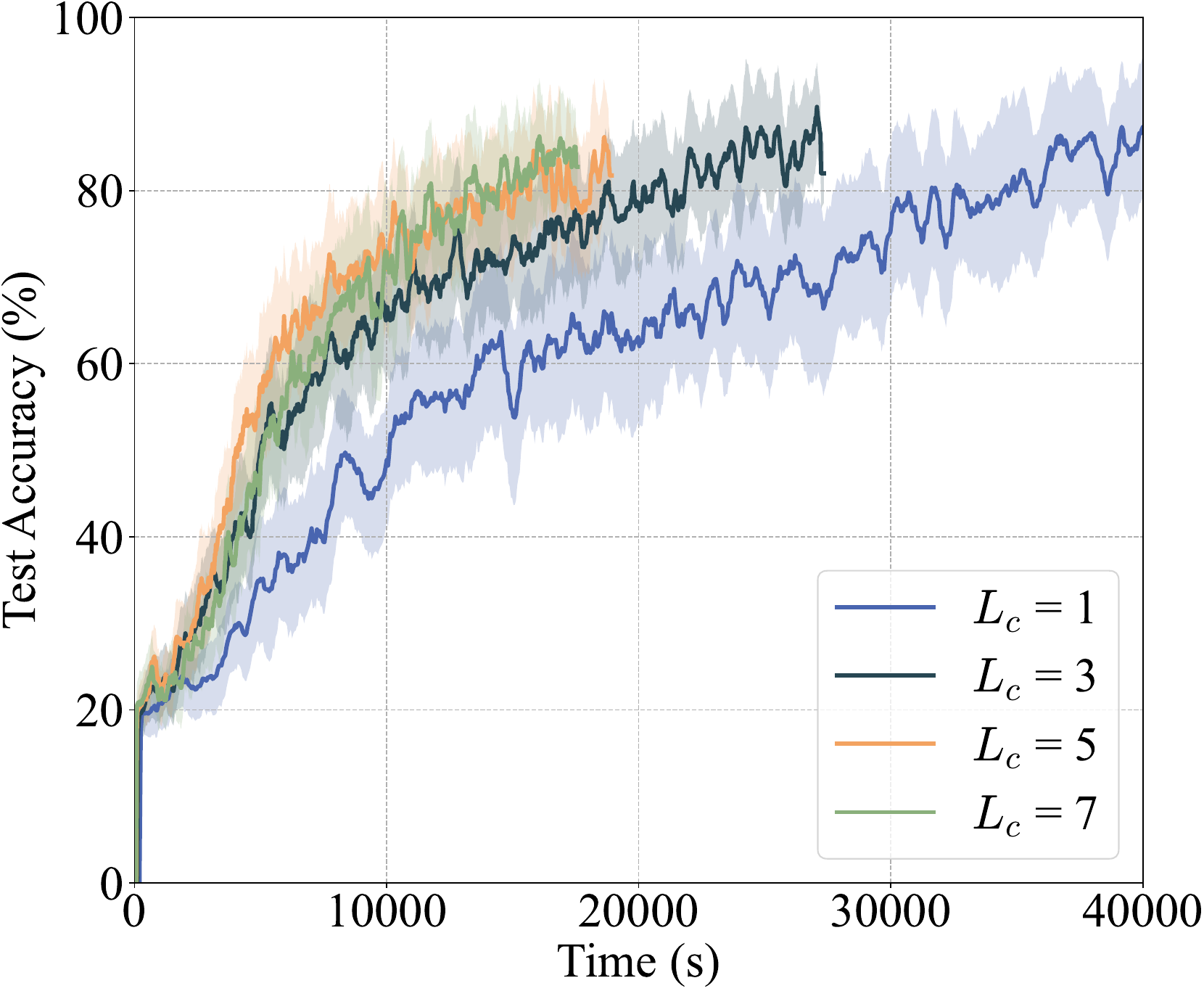}\label{5b}}
	\hspace{2pt}
    \subfigure[SFLSC3 with different $b$.]{
		\includegraphics[width=0.23\linewidth]{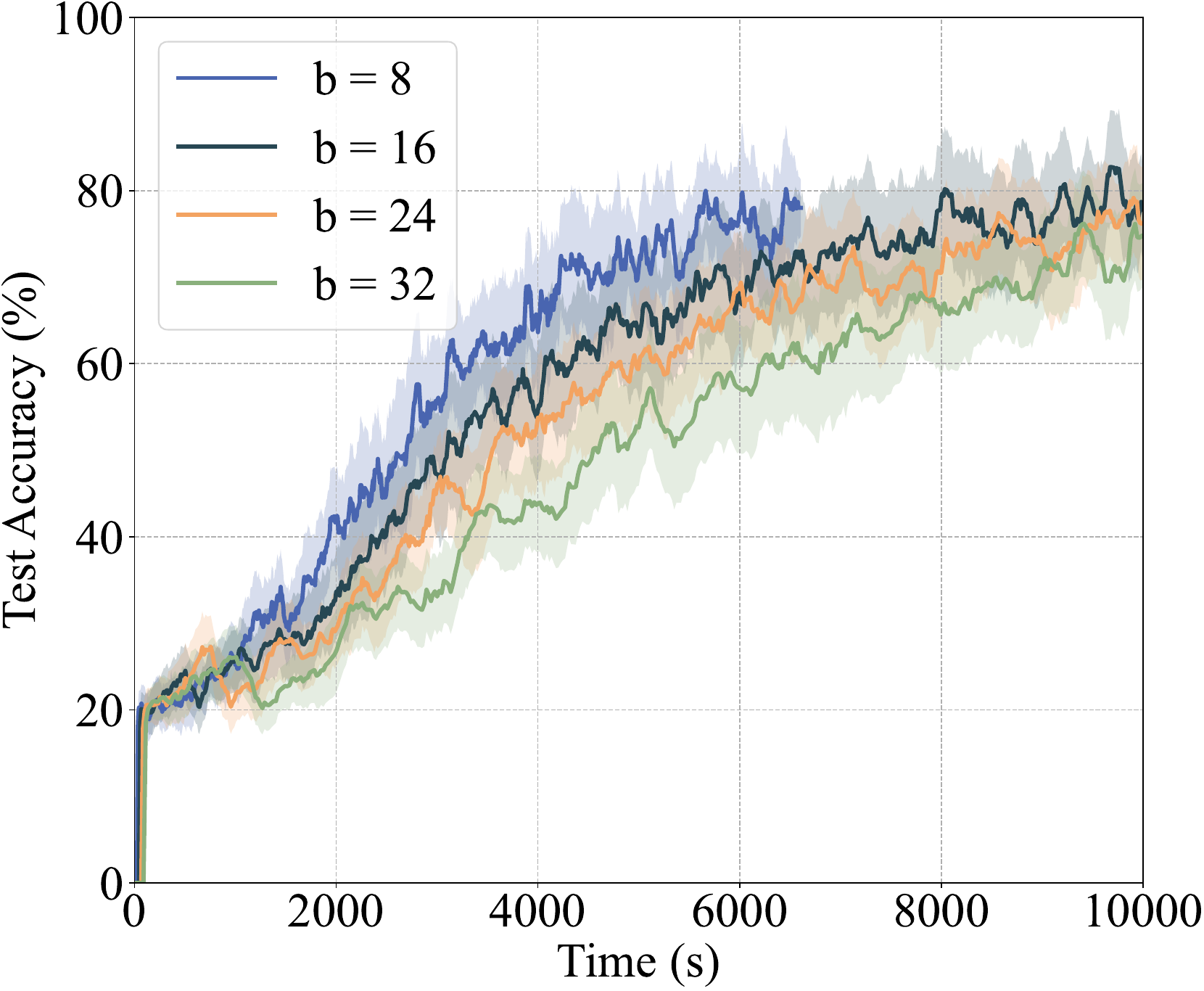}\label{5c}}
	\hspace{2pt}
    \subfigure[SFLSC3 with different $\theta_s$.]{ 
		\includegraphics[width=0.23\linewidth]{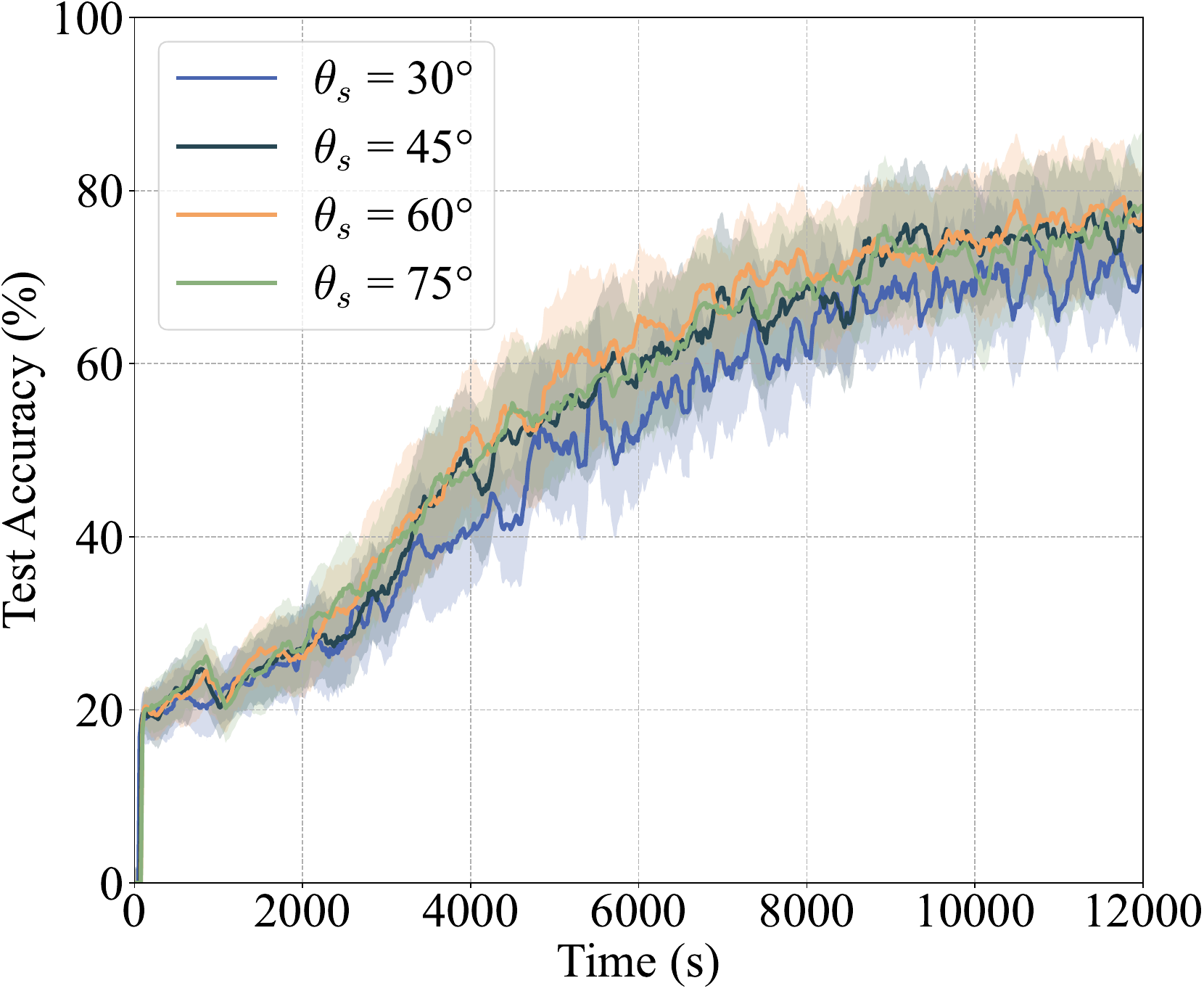}\label{5d}} 
	\caption{The impact of different optimization variables on the convergence vs time of SFLSC3.}
 \label{fig.5}
\end{figure*}

\begin{figure*}[t]
	\centering  
	\subfigbottomskip=1pt 
	\subfigcapskip=-5pt 
	\subfigure[SFLSC3 with different $I$.]{
		\includegraphics[width=0.23\linewidth]{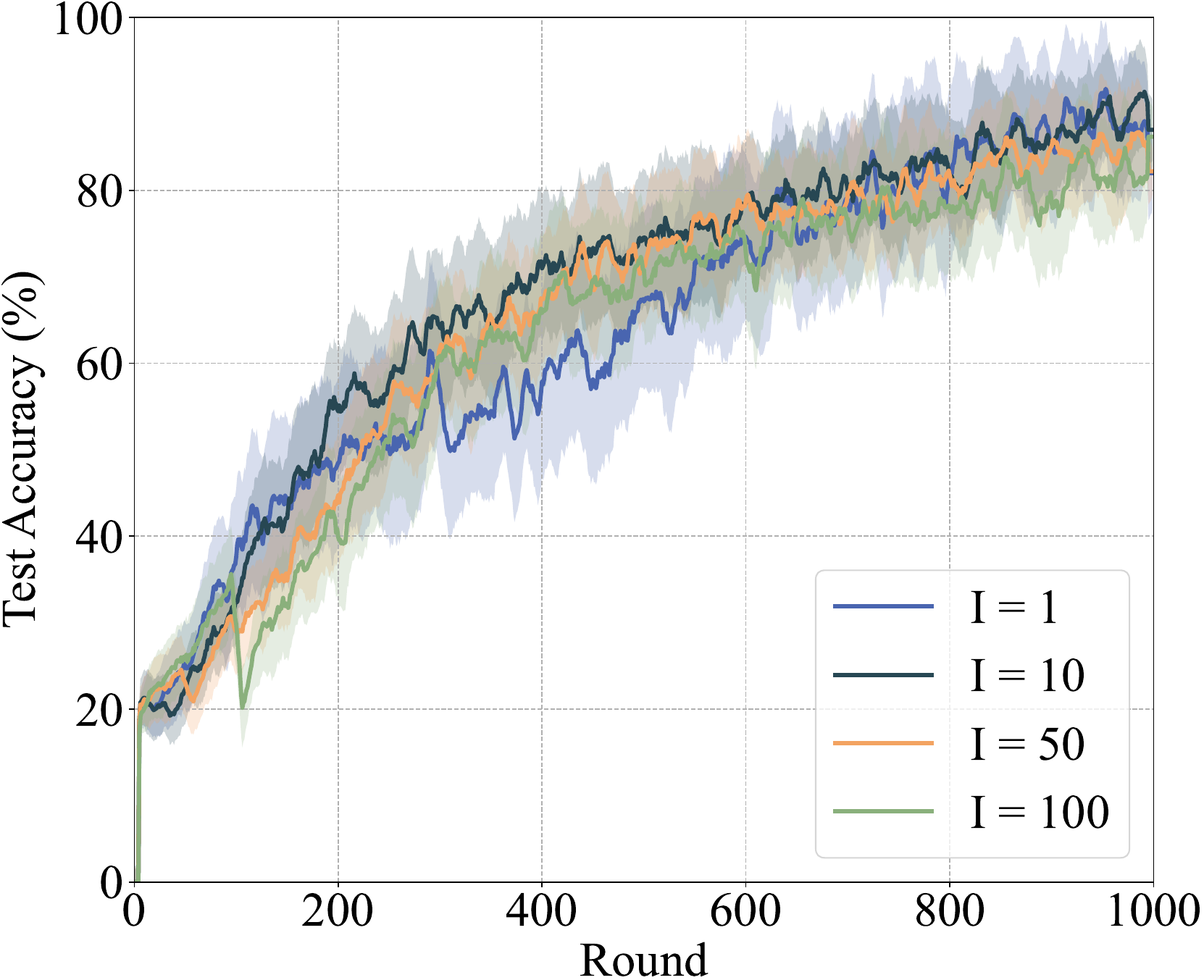}\label{6a}}
	\hspace{2pt}
	\subfigure[SFLSC3 with different $L_c$.]{
		\includegraphics[width=0.23\linewidth]{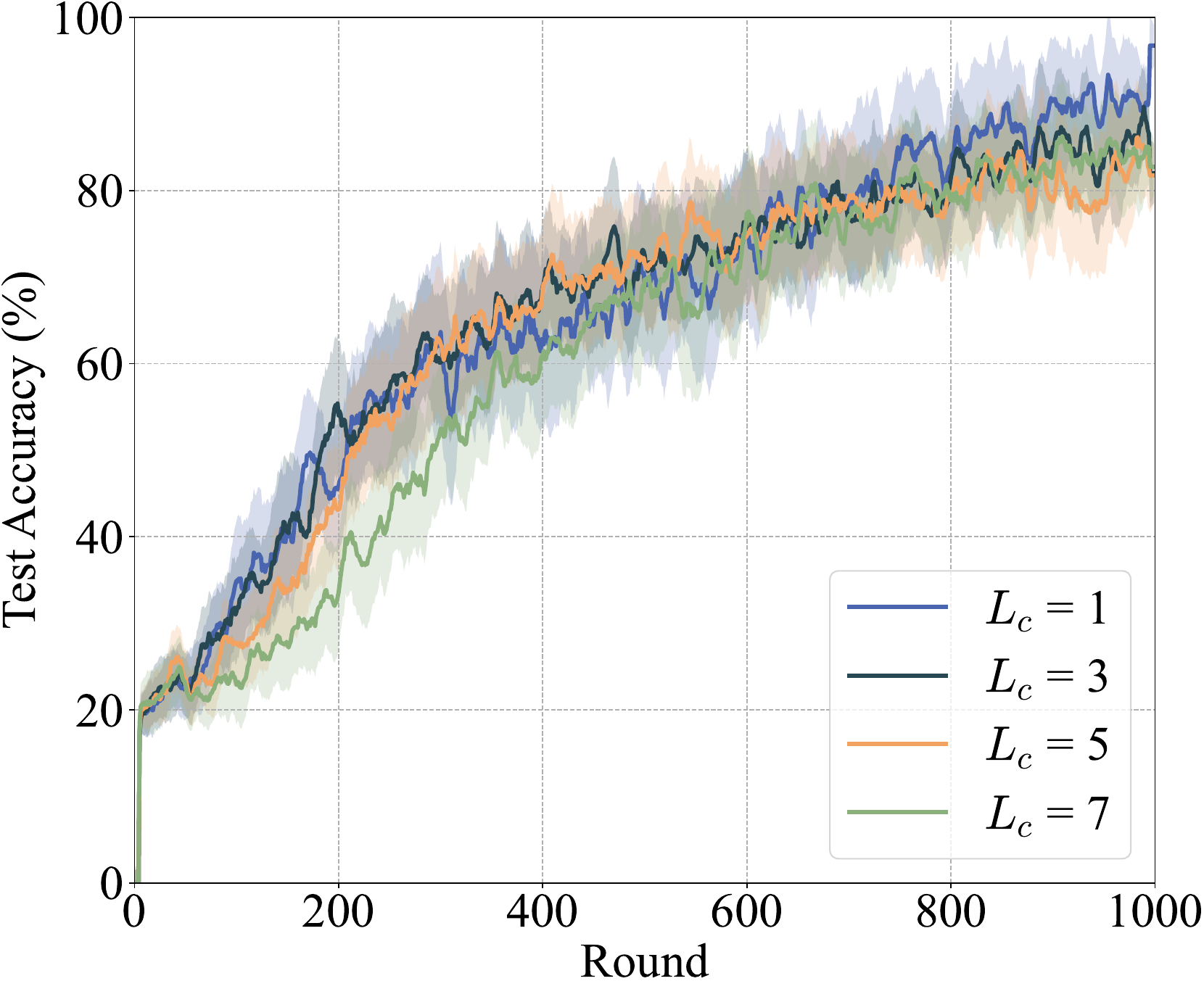}\label{6b}}
	\hspace{2pt}
    \subfigure[SFLSC3 with different $b$.]{
		\includegraphics[width=0.23\linewidth]{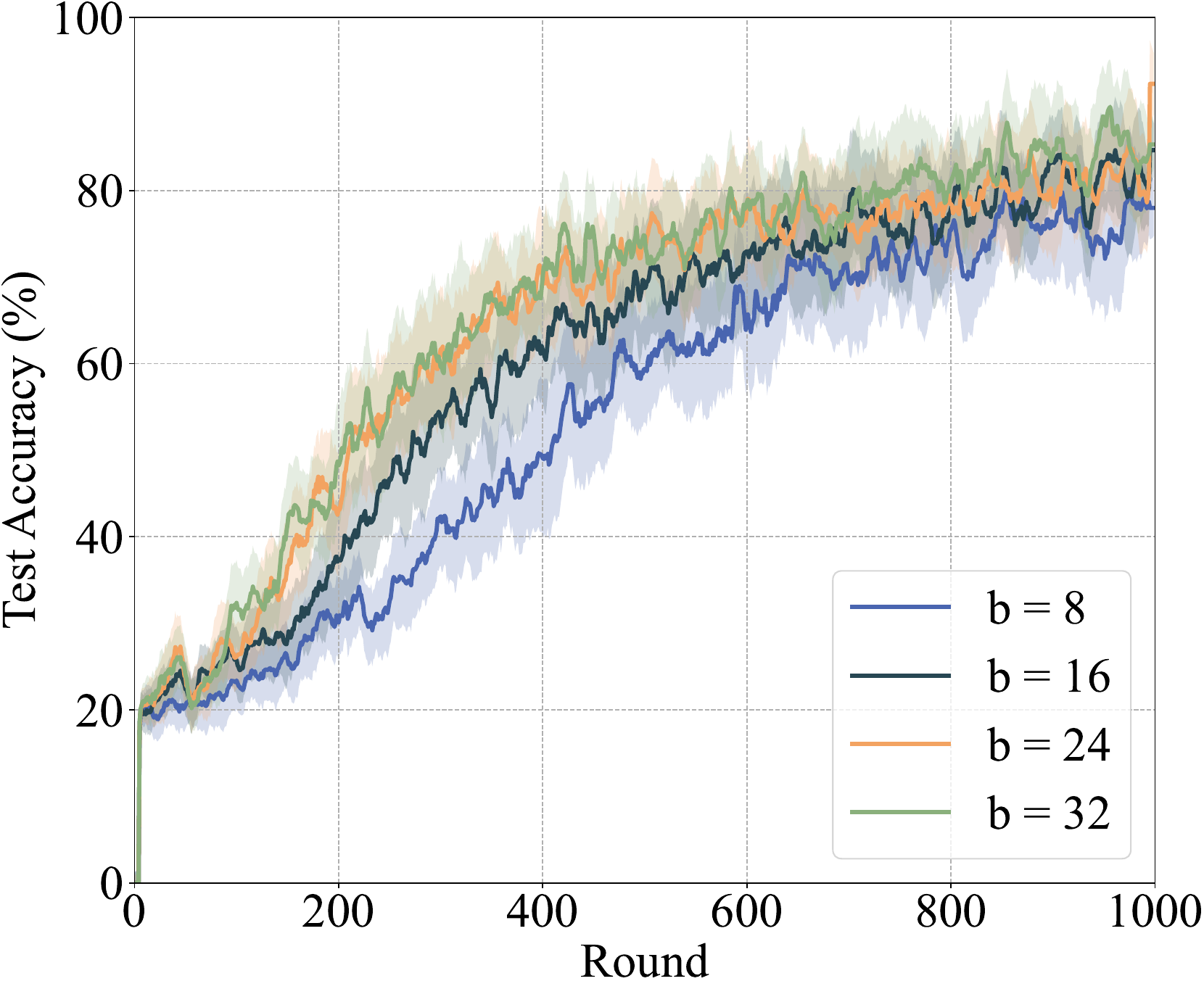}\label{6c}}
	\hspace{2pt}
    \subfigure[SFLSC3 with different $\theta_s$.]{ 
		\includegraphics[width=0.23\linewidth]{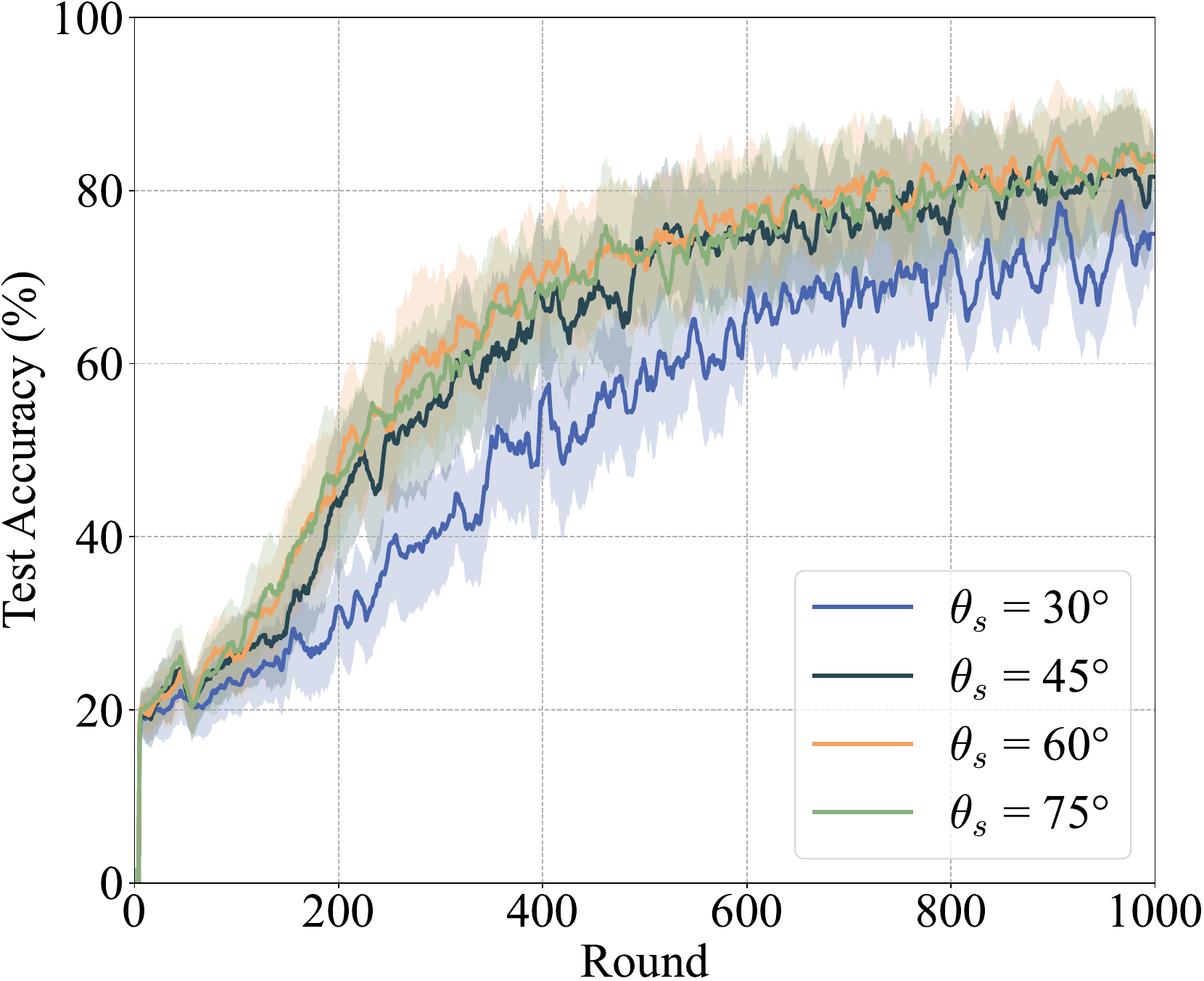}\label{6d}} 
	\caption{The impact of different optimization variables on the convergence vs round of SFLSC3.}
 \label{fig.6}
\end{figure*}
\begin{figure}[t]
	\centering  
	\subfigbottomskip=1pt 
	\subfigcapskip=-5pt 
	\subfigure[Average test loss.]{
		\includegraphics[width=0.8\linewidth]{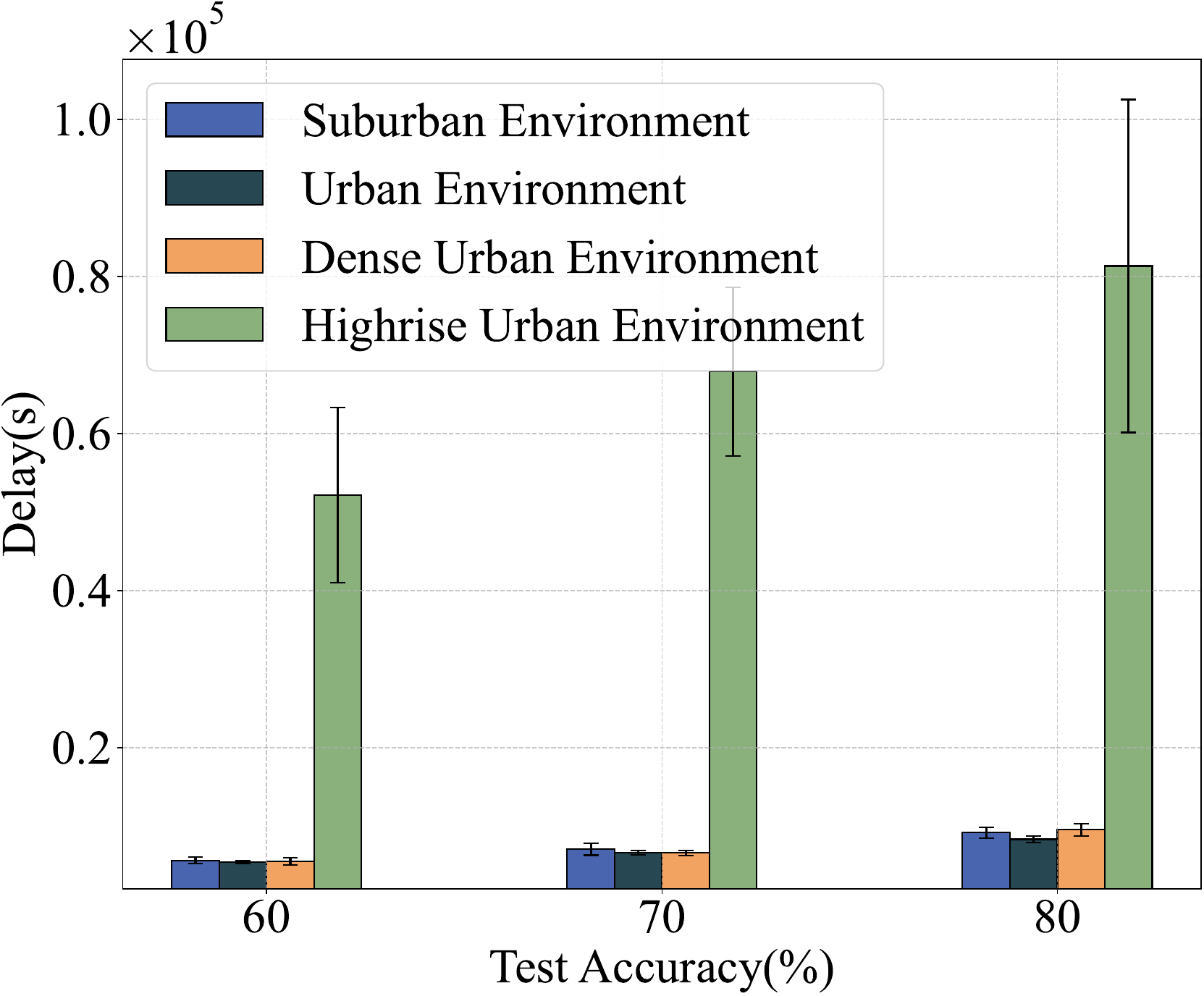}\label{7a}}
        \subfigure[Average test accuracy.]{
		\includegraphics[width=0.8\linewidth]{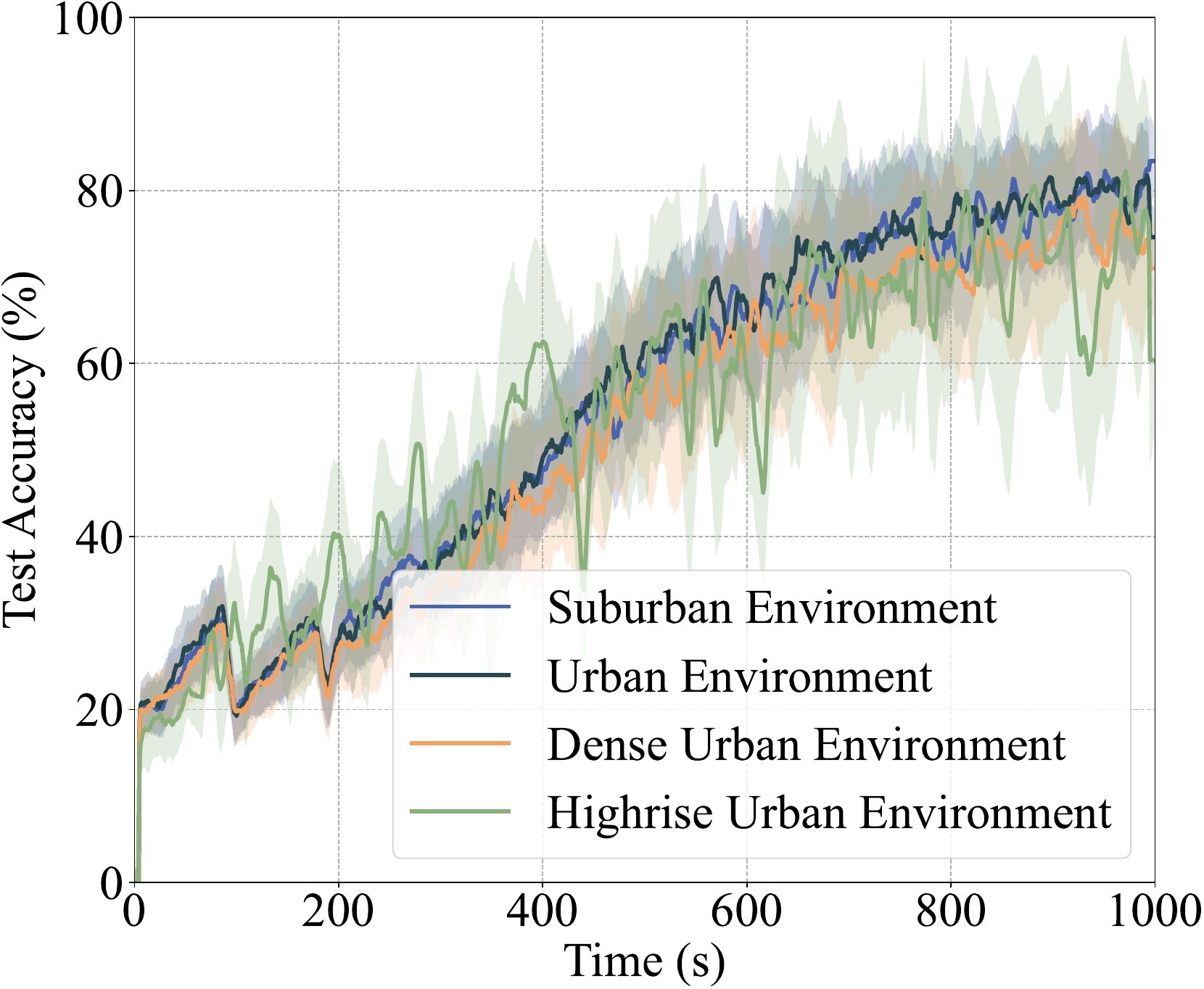}\label{7b}}
	\caption{The impact of different environments on delay and convergence of SFLSC3.}
 \label{fig.7}
\end{figure}
\section{Simulation Results and Discussions}
\subsection{Experiment Settings}
We consider a UAV-assisted ISC3 system with $M$=5 UAVs. We set the sensing transmit power $p_s$=30 dBm, communication transmit power $p_c$=20 dBm, communication bandwidth $B$=1.2 MHz, noise power spectral density $N_0$=-174 dBm/Hz, height $H$=10 m, chirp duration $T_d$=10 $\mu$s, chirp numbers $\lambda$=25, environmental parameters $\varphi$=12.08 and $\varsigma$=0.1139. Then, we use the simulator proposed in\cite{9593198} to generate five different human motion datasets: child/adult standing, child pacing, adult pacing, child walking, and adult walking. The spectrograms corresponding to different human motions can be seen in Fig.~\ref{dataset}. 
We employ ResNet18 as the classification model. The network commences with a large-kernel convolution and downsampling, followed by four groups of residual blocks that utilize pre-activation units to enhance information flow and training efficiency. As the network progresses, it incrementally increases the number of channels while decreasing spatial resolution. High-level features are extracted through a global average pooling layer, and the predicted class is ultimately output by a fully connected layer.

To demonstrate the effectiveness of our proposed SFLSC3, we compared our scheme with the following schemes:

\begin{itemize}
    \item \textbf{Ours: }This scheme is based on the proposed SFLSC3 framework and the control algorithm presented in Section IV, jointly optimizing the UAV positions $\left\{\boldsymbol{u}_m\right\}$, client aggregation frequency $I$, the data sensing volume $b$, and the split point $L_c$.
    \item \textbf{Baseline 1: }This scheme considers that the client aggregation frequency $I$ is given, and jointly optimizes the UAV positions $\left\{\boldsymbol{u}_m\right\}$, the data sensing volume $b$, and the split point $L_c$, as discussed in Section IV.
    \item \textbf{Baseline 2: }This scheme considers that the split point $L_c$ is given, and jointly optimizes the client aggregation frequency $I$, the UAV positions $\left\{\boldsymbol{u}_m\right\}$, the data sensing volume $b$ as discussed in Section IV.
    \item \textbf{Baseline 3: }This scheme considers that the data sensing volume $b$ is set to its maximum, and jointly optimizes the UAV positions $\left\{\boldsymbol{u}_m\right\}$, the client aggregation frequency $I$, and the split point $L_c$, as discussed in Section IV.
    \item \textbf{Baseline 4: }This scheme considers that the UAV positions $\left\{\boldsymbol{u}_m\right\}$ are fixed, and jointly optimizes the client aggregation frequency $I$, the data sensing volume $b$, and the split point $L_c$, as discussed in Section IV.
\end{itemize}
\subsection{Experiment Analysis}
To demonstrate the performance improvement of our proposed optimization algorithm, we compare the convergence and the required delay to achieve the specified model accuracy among different schemes, as shown in Fig.~\ref{fig.3}. It shows that the proposed optimization algorithm significantly outperforms the baselines in terms of total training delay, while achieving comparable convergence speed to baseline 3, which has the highest sensing quality.

As shown in Figs.~\ref{fig.4} and~\ref{fig.5}, we investigate the effects of key decision variables on both the convergence behavior and delay performance of SFLSC3. The results indicate that none of the control factors exhibits a purely monotonic impact on system performance; instead, clear trade-offs can be observed.
First, as illustrated in Figs.~\ref{4a},~\ref{5a} and~\ref{6a}, increasing the client-side aggregation frequency (i.e., decreasing $I$) indeed accelerates model convergence in terms of the number of communication rounds. This improvement can be attributed to the fact that more frequent aggregations effectively reduce the divergence among client-side sub-models. However, each aggregation operation incurs additional communication and computation overhead, leading to a higher average per-round delay. Consequently, when evaluated from a long-term perspective, a smaller $I$ does not necessarily result in a shorter wall-clock time to reach a target accuracy. Instead, an intermediate aggregation interval achieves a more favorable trade-off between convergence speed and overall delay.
Second, the impact of the split point $L_c$ exhibits a similar non-monotonic behavior. As shown in Figs.~\ref{4b}, ~\ref{5b} and~\ref{4b}, a shallower split point (i.e., a smaller $L_c$) leads to faster convergence in terms of rounds, since the majority of the model parameters are maintained and aggregated at the edge server, resulting in more stable updates. Nevertheless, the delay performance does not monotonically improve as $L_c$ decreases. When $L_c$ approaches the boundary values, the delay performance deteriorates. This is because although a smaller $L_c$ reduces the size of the client-side model to be transmitted, it simultaneously increases the volume of intermediate feature data that must be exchanged at the split layer, which may dominate the communication overhead. These observations indicate that the effects of $I$ and $L_c$ on convergence and delay are inherently coupled.
Third, as shown in Figs.~\ref{4c}, ~\ref{5c} and~\ref{6c}, increasing the data sensing volume $b$ accelerates convergence in terms of rounds, owing to the improved data diversity and learning effectiveness at each UAV. However, this comes at the cost of increased sensing, processing, and transmission delay at the UAV side. Interestingly, under the considered experimental settings, a smaller $b$ yields better delay performance, which is mainly due to the relatively simple learning task. Nevertheless, when $b$ becomes excessively small, the convergence performance degrades significantly, revealing another clear trade-off between learning efficiency and delay.
Finally, we examine the influence of UAV positions $\left\{\boldsymbol{u}_m\right\}$, characterized by the sensing angle $\theta_s$. Figs.~\ref{4d}, ~\ref{5d} and~\ref{6d} show that increasing $\theta_s$ improves the sensing success probability, but also leads to higher delay due to the enlarged sensing and communication overhead. Therefore, $\theta_s$ is not always preferable when set to larger values, and an appropriate balance must be achieved. In summary, all the considered control variables, including $I$, $L_c$, $b$, and $\theta_s$, involve inherent trade-offs between convergence efficiency and delay performance. These observations further demonstrate the necessity and practicality of the proposed optimization framework.

The aforementioned experiments are all conducted under relatively ideal conditions. To investigate the impact of different environments on the performance of SFLSC3 and thereby guide subsequent designs, we conduct experiments in suburban, urban, dense urban, and high-rise urban environments, as shown in Fig.~\ref{fig.7}. Complex environments can deteriorate model convergence, as occlusions in such environments affect the probability of Los links, thereby impacting the quality of UAV sensing and communication channels.

\section{Conclusion}
This paper presents a novel SFL framework for UAV-enabled integrated sensing, computation, and communication, named SFLSC3, addressing key challenges in UAV-assisted FEL systems. By leveraging the unique capabilities of UAVs for local data collection and computation, SFLSC3 optimizes the trade-off between training accuracy and delay. Our theoretical analysis provides insights into the convergence behavior of SFLSC3, establishing upper bounds for the convergence gap based on UAV deployment, split point selection, and data sensing volume. We also propose a joint optimization approach to minimize delay while ensuring the desired model accuracy. The non-convex nature of the problem is tackled with a low-complexity algorithm that efficiently delivers near-optimal solutions. Extensive experiments on a target motion recognition task demonstrate the superiority of SFLSC3 in terms of convergence and latency performance compared to existing methods.

\section*{APPENDIX A}
Let's fix the training round at $t \geq 1$. Identify the largest $t_0 \leq t$ and $t_0$ is a multiple of $I$ (i.e. $t_0$ mod $I=0$). It should be noted that such a $t_0$  definitely exists and the difference $t-t_0$ is at most $I$. According to the Eq.\eqref{modelcm} which are used to update the model weights, we have
\begin{equation}
    \boldsymbol{w}_{c,m}^t = \boldsymbol{w}_{c}^{t_0}-\eta\sum\limits_{\tau=t_0}^{t-1}\alpha_{m}^{\tau}\mathbf{g}_{c,m}^{\tau}
\end{equation}
and by Eq.\eqref{modelc}, we have
\begin{equation}
    \boldsymbol{w}_{c}^t = \boldsymbol{w}_{c}^{t_0}-\eta\sum\limits_{\tau=t_0}^{t-1}\frac{1}{M}\sum\limits_{m=1}^M\alpha_{m}^{\tau}\mathbf{g}_{c,m}^{\tau}.
\end{equation}

Thus, we have
\begin{equation}
\begin{aligned}
&\mathbb{E}\left[\left\|\boldsymbol{w}_c^{t}-\boldsymbol{w}_{c,m}^{t}\right\|^2\right] \\
&=\eta^2\mathbb{E}\left[\left\|\sum\limits_{\tau=t_0}^{t-1}\frac{1}{M}\sum\limits_{m=1}^{M}\alpha_m^{\tau}\mathbf{g}_{c,m}^{\tau}-\sum\limits_{\tau=t_0}^{t-1}\alpha_m^{\tau}\mathbf{g}_{c,m}^{\tau}\right\|^2\right] \\
&\overset{\left(a\right)}{\leq}
2\eta^2\mathbb{E}\left[\left\|\sum\limits_{\tau=t_0}^{t-1}\frac{1}{M}\sum\limits_{m=1}^{M}\alpha_m^{\tau}\mathbf{g}_{c,m}^{\tau}\right\|^2+\left\|\sum\limits_{\tau=t_0}^{t-1}\alpha_m^{\tau}\mathbf{g}_{c,m}^{\tau}\right\|^2\right]\\
&\overset{\left(b\right)}{\leq}
2\eta^2\left(t-t_0\right)\mathbb{E}\left[\sum\limits_{\tau=t_0}^{t-1}\left\|\frac{\sum\limits_{m=1}^{M}\alpha_m^{\tau}\mathbf{g}_{c,m}^{\tau}}{M}\right\|^2+\sum\limits_{\tau=t_0}^{t-1}\left\|\alpha_m^{\tau}\mathbf{g}_{c,m}^{\tau}\right\|^2\right]\\
&\overset{\left(c\right)}{\leq}
2\eta^2\left(t-t_0\right)\mathbb{E}\left[\frac{\sum\limits_{\tau=t_0}^{t-1}\sum\limits_{m=1}^{M}\left\|\alpha_m^{\tau}\mathbf{g}_{c,m}^{\tau}\right\|^2}{M}+\sum\limits_{\tau=t_0}^{t-1}\left\|\alpha_m^{\tau}\mathbf{g}_{c,m}^{\tau}\right\|^2\right]\\
&\overset{\left(d\right)}{\leq}
4\eta^2\left(t-t_0\right)^2\sum\limits_{l=1}^{L_c}G_l^2\\
&\leq 4\eta^2I^2\sum\limits_{l=1}^{L_c}G_l^2,
\end{aligned}
\end{equation}
where inequality $\left(a\right)-\left(c\right)$ follows from $\|\sum\limits_{i=1}^nx_i\|^2\leq n\sum\limits_{i=1}^n\|x_i\|^2$, and inequality $\left(d\right)$ is due to \textbf{Assumption 2}.
\section*{APPENDIX B}
To begin with, we derive the expectation expression of $\boldsymbol{w}^t-\boldsymbol{w}^{t-1}$.
\begin{equation}
\begin{aligned}
&\mathbb{E}\left[\frac{\sum\limits_{m=1}^M\alpha_m^{t-1}\mathbf{g}_m^{t-1}}{M}\Bigg|\sum\limits_{m=1}^M\alpha_m^{t-1}\neq 0\right]\\&=\mathbb{E}\left[\sum\limits_{l=1}^M\sum\limits_{\substack{|\mathcal{M}_1|=l\\|\mathcal{M}_2|=M-l}}^{\mathcal{M}_1\cup \mathcal{M}_2=\mathcal{M}} Pr\Bigg(\alpha_{m_1}^{t-1} = 1 \forall m_1 \in \mathcal{M}_1,\alpha_{m_2}^{t-1} = 0 \right.  \\
& \left. \quad \forall m_2 \in \mathcal{M}_2 \Bigg|\sum\limits_{m=1}^M\alpha_m^{t-1}\neq0\Bigg)\frac{1}{M}\sum\limits_{m_1 \in \mathcal{M}_1}\mathbf{g}_{m_1}^{t-1}\right]\\&= \mathbb{E}\left[\sum\limits_{l=1}^M\sum\limits_{\substack{|\mathcal{M}_1|=l\\|\mathcal{M}_2|=M-l}}^{\mathcal{M}_1\cup \mathcal{M}_2=\mathcal{M}}\frac{1}{M}\sum\limits_{m_1 \in \mathcal{M}_1}\mathbf{g}_{m_1}^{t-1} \right.  \\
& \left.\quad\cdot\frac{\prod\limits_{m_1 \in \mathcal{M}_1}q_{s,m_1}\prod\limits_{m_2 \in \mathcal{M}_2}(1-q_{s,m_2})}{1-\prod\limits_{m \in M}(1-q_{s,m})}\right]\\&
\overset{\triangle}{=}\sum\limits_{m=1}^M\varphi_m\mathbb{E}\left[\mathbf{g}_m^{t-1}\right],
\end{aligned}
\end{equation}
where $\mathcal{M}_1$ represents the set of UAVs which succeed in sensing the target, while $\mathcal{M}_2$ denotes the set of UAVs which fail to do that. Besides, $\varphi_m, \forall m \in \mathcal{M}$ is related to $q_{s,m}$ and $\sum\limits_{m=1}^M\varphi_m\leq1$, this can be observed by setting $\mathbf{g}_m^{t-1}=1$.

Next, we prove \textbf{Lemma 2} as follows
\begin{equation}
\begin{aligned}
&\mathbb{E}\left[\langle \nabla \mathcal{L}\left(\boldsymbol{w}^{t-1}\right),\boldsymbol{w}^t-\boldsymbol{w}^{t-1}\rangle\right]\\
&= -\eta\mathbb{E}\left[\left\langle \nabla \mathcal{L}\left(\boldsymbol{w}^{t-1}\right), \frac{\sum\limits_{m=1}^M \alpha_m^{t-1} \mathbf{g}_m^{t-1}}{M}\right\rangle\right]\\
&=
-\eta \mathbb{E}\left[\left\langle \nabla\mathcal{L}\left(\boldsymbol{w}^{t-1}\right), 
\sum\limits_{m=1}^M \varphi_m \nabla \mathcal{L}_m\left(\boldsymbol{w}_m^{t-1}\right)\right\rangle\right]\\
&\leq
-\eta \mathbb{E}\left[\left\langle \sum\limits_{m=1}^M \varphi_m \nabla \mathcal{L}\left(\boldsymbol{w}^{t-1}\right), 
\sum\limits_{m=1}^M \varphi_m \nabla \mathcal{L}_m\left(\boldsymbol{w}_m^{t-1}\right)\right\rangle\right]\\
&\overset{\left(a\right)}{=} \frac{\eta}{2} \mathbb{E}\left[\left\|\sum\limits_{m=1}^M \varphi_m \left(\nabla \mathcal{L}\left(\boldsymbol{w}^{t-1}\right) - \nabla \mathcal{L}_m\left(\boldsymbol{w}_m^{t-1}\right)\right)\right\|^2 \right.\\
&\quad \left.- \left\|\sum\limits_{m=1}^M \varphi_m \nabla \mathcal{L}\left(\boldsymbol{w}^{t-1}\right)\right\|^2 - 
\left\|\sum\limits_{m=1}^M \varphi_m \nabla \mathcal{L}_m\left(\boldsymbol{w}_m^{t-1}\right)\right\|^2 \right] \\
&\leq \underbrace{\frac{\eta}{2} \mathbb{E}\left[\left\|\sum\limits_{m=1}^M \varphi_m \left(\nabla \mathcal{L}\left(\boldsymbol{w}^{t-1}\right) - \nabla \mathcal{L}_m\left(\boldsymbol{w}_m^{t-1}\right)\right)\right\|^2\right]}_{\overset{\triangle}{=}X_B}\\
&\quad  - \frac{\eta}{2}\mathbb{E}\left[\left\|\sum\limits_{m=1}^M \varphi_m \nabla \mathcal{L}\left(\boldsymbol{w}^{t-1}\right)\right\|^2 \right],
\end{aligned}
\end{equation}
where equality $\left(a\right)$ follows from $\langle \boldsymbol{x},\boldsymbol{y}\rangle=\frac{1}{2}(\|\boldsymbol{x}\|^2+\|\boldsymbol{y}\|^2-\|\boldsymbol{x}-\boldsymbol{y}\|^2)$. Then, we derive the upper bound of $X_B$.
\begin{equation}
\begin{aligned}
&X_B= \frac{\eta}{2} \mathbb{E}\left[ \displaystyle\left\|\sum\limits_{m=1}^M \varphi_m\left(\nabla \mathcal{L}\left(\boldsymbol{w}^{t-1}\right) - \nabla \mathcal{L}_m\left(\boldsymbol{w}^{t-1}\right) \right. \right. \right. \\
&\quad \left. \left. \left. + \nabla \mathcal{L}_m\left(\boldsymbol{w}^{t-1}\right)- \nabla\mathcal{L}_m\left(\boldsymbol{w}^{t-1}_m\right) \right)\vphantom{\sum\limits_{m=1}^M}\right\|^2 \right] \\
&\overset{\left(b\right)}{\leq} \eta \mathbb{E}\left[ \displaystyle\left\|\sum\limits_{m=1}^M \varphi_m\left(\nabla \mathcal{L}\left(\boldsymbol{w}^{t-1}\right) - \nabla \mathcal{L}_m\left(\boldsymbol{w}^{t-1}\right)\right)\right\|^2  \right.\\
&\quad \left. + \left\|\sum\limits_{m=1}^M \varphi_m\left(\nabla \mathcal{L}_m\left(\boldsymbol{w}^{t-1}\right) - \nabla\mathcal{L}_m\left(\boldsymbol{w}^{t-1}_m\right)\right) \vphantom{\sum\limits_{m=1}^M}\right\|^2 \right] \\
&\overset{\left(c\right)}{\leq}
\eta\sum\limits_{m=1}^M \varphi_m^2 \mathbb{E}\left[ \displaystyle\sum\limits_{m=1}^M\left\|\nabla \mathcal{L}\left(\boldsymbol{w}^{t-1}\right) - \nabla \mathcal{L}_m\left(\boldsymbol{w}^{t-1}\right)\right\|^2  \right.\\
&\quad \left. + \sum\limits_{m=1}^M \left\|\nabla \mathcal{L}_m\left(\boldsymbol{w}^{t-1}\right) - \nabla\mathcal{L}_m\left(\boldsymbol{w}^{t-1}_m\right) \right\|^2  \right]\\
&- \quad \frac{\eta}{2} \mathbb{E}\left[\left\|\sum\limits_{m=1}^M \varphi_m \nabla \mathcal{L}\left(\boldsymbol{w}^{t-1}\right)\right\|^2\right]
\end{aligned}
\end{equation}

\begin{equation*}
\begin{aligned}
&\overset{\left(d\right)}{\leq}
\eta  \left(\sum\limits_{m=1}^M \Lambda_m^2+\beta^2\sum\limits_{m=1}^M\mathbb{E}\left[\left\|\boldsymbol{w}^{t-1}-\boldsymbol{w}^{t-1}_m\right\|^2\right]\right) \\
&\leq
\eta \sum\limits_{m=1}^M \Lambda_m^2+\eta\beta^2\sum\limits_{m=1}^M\mathbb{E}\left[\|\boldsymbol{w}^{t-1}_{s}-\boldsymbol{w}^{t-1}_{s,m}\|^2 \right.\\
&\left.+\|\boldsymbol{w}^{t-1}_{c}-\boldsymbol{w}^{t-1}_{c,m}\|^2\right]\\
&\leq
  \sum\limits_{m=1}^M \eta\Lambda_m^2 +4M\beta^2\eta^3I^2\sum\limits_{l=1}^{L_c}G_l^2,
\end{aligned}
\end{equation*}
where inequality $\left(b\right)$ is due to $\|\boldsymbol{x}+y\|^2\leq2\|\boldsymbol{x}\|^2+2\|\boldsymbol{y}\|^2$ and inequality $\left(c\right)$ stems from $\|\sum\limits_{i=1}^N\boldsymbol{x}_i\boldsymbol{y}_i\|^2\leq\sum\limits_{i=1}^N\|\boldsymbol{x}_i\|^2+\sum\limits_{i=1}^N\|\boldsymbol{y}_i\|^2$, inequality $\left(d\right)$ results from \textbf{Assumption } 1 and 3.

Thus, we can derive that
\begin{equation}
\begin{aligned}
&\mathbb{E}\left[\langle \nabla \mathcal{L}\left(\boldsymbol{w}^{t-1}\right),\boldsymbol{w}^t-\boldsymbol{w}^{t-1}\rangle\right]\leq \sum\limits_{m=1}^M \eta\Lambda_m^2 \\&+4M\beta^2\eta^3I^2\sum\limits_{l=1}^{L_c}G_l^2- \frac{\eta}{2}\mathbb{E}\left[\left\|\sum\limits_{m=1}^M \varphi_m \nabla \mathcal{L}\left(\boldsymbol{w}^{t-1}\right)\right\|^2 \right].
\end{aligned}
\end{equation}
\section*{APPENDIX C}
\begin{equation}
\begin{aligned}
& \mathbb{E}\left[\|\boldsymbol{w}^t-\boldsymbol{w}^{t-1}\|^2\right] = \eta^2 \mathbb{E}\left[\left\|\frac{\sum\limits_{m=1}^M \alpha_m^{t-1} \mathbf{g}_m^{t-1}}{M}\right\|^2\right] \\
& \quad = \eta^2 \mathbb{E}\left[\left\|\frac{\sum\limits_{M=1}^m \alpha_m^{t-1} \left(\mathbf{g}_m^{t-1} - \nabla \mathcal{L}_m(\boldsymbol{w}_m^{t-1})\right)}{M} \right. \right. \\
& \quad \left. \left. + \frac{\sum\limits_{m=1}^M \alpha_m^{t-1}  \nabla \mathcal{L}_m(\boldsymbol{w}^{t-1}_m)}{M}\right\|^2 \right]\\
&\leq 2\eta^2\underbrace{\mathbb{E}\left[\left\|\frac{\sum\limits_{m=1}^M \alpha_m^{t-1} \left(\mathbf{g}_m^{t-1} - \nabla \mathcal{L}_m(\boldsymbol{w}_m^{t-1})\right)}{M}\right\|^2\right]}_{\triangleq X}\\
&+2\eta^2\underbrace{\mathbb{E}\left[\left\|\frac{\sum\limits_{m=1}^M \alpha_m^{t-1}  \nabla \mathcal{L}_m(\boldsymbol{w}^{t-1}_m)}{M}\right\|^2\right]}_{\triangleq Y},
\end{aligned}
\end{equation}
Then, we derive the upper bound of $X$ and $Y$, respectively.

\subsubsection{Bound of $X$}
\begin{equation}
\begin{aligned}
X&\overset{\left(a\right)}{=}
\mathbb{E}\left[\frac{\sum\limits_{m=1}^M \alpha_m^{t-1} \left\|\left(\mathbf{g}_m^{t-1} - \nabla \mathcal{L}_m(\boldsymbol{w}_m^{t-1})\right)\right\|^2}{M^2}\right]\\&
+\mathbb{E}\left[\frac{\sum\limits_{i=1}^M \alpha_i^{t-1}\sum\limits_{j=1,j\neq i}^M \alpha_j^{t-1} \triangle_i\triangle_j}{M^2}\right]\\&\leq
\mathbb{E}\left[\frac{\sum\limits_{m=1}^M  \left\|\left(\mathbf{g}_m^{t-1} - \nabla \mathcal{L}_m(\boldsymbol{w}_m^{t-1})\right)\right\|^2+\sum\limits_{i=1}^M \sum\limits_{\substack{j=1\\j\neq i}}^M \triangle_i\triangle_j}{M^2}\right]\\
&\overset{\left(b\right)}{=}\frac{1}{M^2}\sum\limits_{m=1}^M \mathbb{E}\left[\left\|\left(\mathbf{g}_m^{t-1} - \nabla \mathcal{L}_m(\boldsymbol{w}_m^{t-1})\right)\right\|^2\right]\\
&\overset{\left(c\right)}{\leq}\frac{1}{M} \frac{\sum\limits_{l=1}^{L} \sigma_l^2}{b},
\end{aligned}
\label{X}
\end{equation}
where $\left(a\right)$ is due to $\|\boldsymbol{x}_1+\boldsymbol{x}_2+\dots+\boldsymbol{x}_m\|^2=\|\boldsymbol{x}_1\|^2+\|\boldsymbol{x}_2\|^2+\dots+\|\boldsymbol{x}_m\|^2+\sum\limits_{i=1}^m\sum\limits_{j=1,i\neq j}^m \boldsymbol{x}_i\boldsymbol{x}_j$ and $\triangle_i =\mathbf{g}_i^{t-1} - \nabla \mathcal{L}_i(\boldsymbol{w}_i^{t-1}) $, inequality $\left(b\right)$ stems from \textbf{Assumption 4} and inequality $\left(c\right)$ stems from \textbf{Assumption 2}.

\subsubsection{Bound of $Y$}
\begin{equation}
\begin{aligned}
%
Y&\overset{\left(a\right)}{\leq}2\mathbb{E}\left[\frac{\left\|\sum\limits_{m=1}^M \alpha_m^{t-1} \left( \nabla \mathcal{L}_m(\boldsymbol{w}^{t-1}_m)-\nabla\mathcal{L}(\boldsymbol{w}^{t-1}_m)\right)\right\|}{M^2}^2\right]\\&+2\mathbb{E}\left[\frac{\left\|\sum\limits_{m=1}^M \alpha_m^{t-1} \nabla\mathcal{L}(\boldsymbol{w}^{t-1}_m)\right\|}{M^2}^2\right]\\& \overset{\left(b\right)}{=}2\underbrace{\mathbb{E}\left[\frac{\sum\limits_{m=1}^M \alpha_m^{t-1} \left\|\left( \nabla \mathcal{L}_m(\boldsymbol{w}^{t-1}_m)-\nabla\mathcal{L}(\boldsymbol{w}^{t-1}_m)\right)\right\|}{M^2}^2\right]}_{Y_1}\\&+2\underbrace{\mathbb{E}\left[\frac{\sum\limits_{i=1}^M \sum\limits_{j=1,j\neq i}^M\alpha_m^{t-1} \Phi_i \Phi_j}{M^2}^2\right]}_{Y_2}\\
\end{aligned}
\end{equation}
\begin{equation*}
\begin{aligned}
&+2\underbrace{\mathbb{E}\left[\frac{\left\|\sum\limits_{m=1}^M \alpha_m^{t-1} \left(\nabla\mathcal{L}(\boldsymbol{w}^{t-1}_m)\right)\right\|}{M^2}^2\right]}_{Y_3}, 
\end{aligned}
\end{equation*}
where $\left(a\right)$ follows from $\|\boldsymbol{x}+\boldsymbol{y}\|^2\leq2\|\boldsymbol{x}\|^2+2\|\boldsymbol{y}\|^2$, equality $\left(b\right)$ is similar to Eq. \ref{X}$\left(a\right)$ and $\Phi_i = \nabla \mathcal{L}_i(\boldsymbol{w}^{t-1}_i)-\nabla\mathcal{L}(\boldsymbol{w}^{t-1}_i)$. Next, we find the upper bound of $Y_1$, $Y_2$ and $Y_3$ as follows.

Firstly, according to \textbf{Assumption 3}, we have
\begin{equation}
\begin{aligned}
Y_1 &\leq \mathbb{E}\left[\frac{\sum\limits_{m=1}^M \left\|\left( \nabla \mathcal{L}_m(\boldsymbol{w}^{t-1}_m)-\nabla\mathcal{L}(\boldsymbol{w}^{t-1}_m)\right)\right\|}{M^2}^2\right]\\&
\leq \frac{1}{M^2} \sum\limits_{m=1}^M \Lambda_m^2.
\end{aligned}
\end{equation}

Secondly, the upper bound of $Y_2$ can be derived as follows

\begin{equation}
\begin{aligned}
Y_2 &= \mathbb{E}\left[\sum\limits_{l=2}^M\sum\limits_{\substack{|M_1|=l\\|M_2|=M-l}}^{M_1\cup M_2=M} Pr\Bigg(\alpha_{m_1}^{t-1} = 1 \forall m_1 \in M_1,\alpha_{m_2}^{t-1} = 0 \right.  \\
& \left. \quad \forall m_2 \in M_2 \Bigg|\sum\limits_{m=1}^M\alpha_m^{t-1}\geq2\Bigg)\frac{1}{M^2}\sum\limits_{i=1}^{M_1}\sum\limits_{\substack{j=1,j\neq i}}^{M_1}\Phi_i\Phi_j\right]\\&
\leq \mathbb{E}\left[\sum\limits_{l=2}^M\sum\limits_{\substack{|M_1|=l\\|M_2|=M-l}}^{M_1\cup M_2=M} Pr\Bigg(\alpha_{m_1}^{t-1} = 1 \forall m_1 \in M_1,\alpha_{m_2}^{t-1} = 0 \right.  \\
& \left. \quad \forall m_2 \in M_2 \Bigg|\sum\limits_{m=1}^M\alpha_m^{t-1}\geq2\Bigg)\frac{1}{l^2}\sum\limits_{i=1}^{M_1}\sum\limits_{\substack{j=1,j\neq i}}^{M_1}\Phi_i\Phi_j\right]\\&
= \mathbb{E}\left[\sum\limits_{l=2}^M\sum\limits_{\substack{|M_1|=l\\|M_2|=M-l}}^{M_1\cup M_2=M}\frac{1}{l^2}\sum\limits_{i=1}^{M_1}\sum\limits_{\substack{j=1,j\neq i}}^{M_1}\Phi_i\Phi_j \right.  \\
& \left.\quad\cdot\frac{\prod\limits_{m_1 \in M_1}q_{s,m_1}\prod\limits_{m_2 \in M_2}(1-q_{s,m_2})}{1-\prod\limits_{m \in M}(1-q_{s,m})-\sum\limits_{m \in M}\prod\limits_{\substack{m' \in M \\ m' \neq m}}(1-q_{s,k'})}\right]\\&
\leq \sum\limits_{l=2}^M\sum\limits_{\substack{|M_1|=l\\|M_2|=M-l}}^{M_1\cup M_2=M}\mathbb{E}\left[\sum\limits_{i=1}^{M_1}\sum\limits_{\substack{j=1,j\neq i}}^{M_1}q_{s,i}\Phi_iq_{s,j}\Phi_j\right]\cdot \frac{1}{l^2} \\&
\quad \cdot \frac{\prod\limits_{m_2 \in M_2}(1 - q_{s,m_2})}{1 - \prod\limits_{m \in M}(1 - q_{s,m}) - \sum\limits_{m \in M} \prod\limits_{\substack{m' \in M \\ m' \neq m}} \left( 1 - q_{s,m'} \right)} \\&
\overset{\left(a\right)}{<}\sum_{m=1}^M\kappa_m\left((q_{s,m}-\bar{q}_s)^2+\bar{q}_s^2\right)\Lambda_m^2,
\end{aligned}
\end{equation}
where inequality $(a)$ is referred to ~\cite{ISAC-2306}. Finally, we derive the upper bound of $Y_3$ as follows
\begin{equation}
\begin{aligned}
Y_3 &\leq 2\mathbb{E}\left[\frac{\left\|\sum\limits_{m=1}^M\alpha_m^{t-1}\left(\nabla \mathcal{L}(\boldsymbol{w}^{t-1}_m)-\nabla\mathcal{L}(\boldsymbol{w}^{t-1})\right)\right\|^2}{M^2}\right]\\&
\quad + 2\mathbb{E}\left[\frac{\left\|\sum\limits_{m=1}^M\alpha_m^{t-1}\nabla\mathcal{L}(\boldsymbol{w}^{t-1})\right\|^2}{M^2}\right]\\&
\leq 2\mathbb{E}\left[\frac{\sum\limits_{m=1}^M\left(\alpha_m^{t-1}\right)^2\sum\limits_{m=1}^M\left\|\left(\nabla \mathcal{L}(\boldsymbol{w}^{t-1}_m)-\nabla\mathcal{L}(\boldsymbol{w}^{t-1})\right)\right\|^2}{M^2}\right]\\&
\quad + 2\mathbb{E}\left[\frac{\sum\limits_{m=1}^M\left(\alpha_m^{t-1}\right)^2\sum\limits_{m=1}^M\left\|\nabla\mathcal{L}(\boldsymbol{w}^{t-1})\right\|^2}{M^2}\right]\\&
\leq 
2\mathbb{E}\left[\frac{\sum\limits_{m=1}^M\left\|\left(\nabla \mathcal{L}(\boldsymbol{w}^{t-1}_m)-\nabla\mathcal{L}(\boldsymbol{w}^{t-1})\right)\right\|^2}{M}\right]\\&
\quad + 2\mathbb{E}\left[\frac{\sum\limits_{m=1}^M\left\|\nabla\mathcal{L}(\boldsymbol{w}^{t-1})\right\|^2}{M}\right]\\&
\leq 
2\beta^2\mathbb{E}\left[\frac{\sum\limits_{m=1}^M\|\boldsymbol{w}^{t-1}-\boldsymbol{w}^{t-1}_m\|^2}{M}\right]+2\mathbb{E}\left[\|\nabla\mathcal{L}(\boldsymbol{w}^{t-1})\|^2\right]\\&
\leq
8\beta^2\eta^2I^2\sum\limits_{l=1}^{L_c}G_l^2+2\mathbb{E}\left[\|\nabla\mathcal{L}(\boldsymbol{w}^{t-1})\|^2\right].
\end{aligned}
\end{equation}

Therefore, the upper bound of Y can be represented by
\begin{equation}
\begin{aligned}
Y &< \frac{2}{M^2} \sum\limits_{m=1}^M \Lambda_m^2+2\sum_{m=1}^M\kappa_m\left((q_{s,m}-\bar{q}_s)^2+\bar{q}_s^2\right)\Lambda_m^2\\&\quad+16\beta^2\eta^2I^2\sum\limits_{l=1}^{L_c}G_l^2+4\mathbb{E}\left[\|\nabla\mathcal{L}(\boldsymbol{w}^{t-1})\|^2\right].
\end{aligned}
\end{equation}

Add X to Y, and we can obtain
\begin{equation}
\begin{aligned}
&\mathbb{E}\left[\|\boldsymbol{w}^t-\boldsymbol{w}^{t-1}\|^2\right] < \frac{\sum\limits_{l=1}^{L} 2\eta^2\sigma_l^2}{Mb}+\frac{4\eta^2}{M^2} \sum\limits_{m=1}^M \Lambda_m^2\\&+4\eta^2\sum_{m=1}^M\kappa_m\left((q_{s,m}-\bar{q}_s)^2+\bar{q}_s^2\right)\Lambda_m^2\\&+32\beta^2\eta^4I^2\sum\limits_{l=1}^{L_c}G_l^2+8\eta^2\mathbb{E}\left[\|\nabla\mathcal{L}(\boldsymbol{w}^{t-1})\|^2\right].
\end{aligned}
\end{equation}

\section*{APPENDIX D}
Based on the smoothness of the loss function $\mathcal{L}(\cdot)$, for any training round $t\geq0$, the second-order Taylor expansion of $\mathcal{L}(\cdot)$ can be expressed as
\begin{equation}
\begin{aligned}
\mathbb{E}\left[\mathcal{L}\left(\boldsymbol{w}^t\right)\right]&\leq \mathbb{E}\left[\mathcal{L}\left(\boldsymbol{w}^{t-1}\right)\right]+\frac{\beta}{2}\mathbb{E}\left[\|\boldsymbol{w}^t-\boldsymbol{w}^{t-1}\|^2\right]\\&\quad+\mathbb{E}\left[\langle \nabla \mathcal{L}\left(\boldsymbol{w}^{t-1}\right),\boldsymbol{w}^t-\boldsymbol{w}^{t-1}\rangle\right].
\end{aligned}
\end{equation}

And according to \textbf{Lemma 2} and \textbf{Lemma 3}, we have
\begin{equation}
\begin{aligned}
\mathbb{E}\left[\mathcal{L}\left(\boldsymbol{w}^t\right)\right]&\leq\mathbb{E}\left[\mathcal{L}\left(\boldsymbol{w}^{t-1}\right)\right]+\frac{\sum\limits_{l=1}^{L_c} \beta\eta^2\sigma_l^2}{Mb}\\&\quad +\frac{2\beta\eta^2}{M^2} \sum\limits_{m=1}^M \Lambda_m^2+16\beta^3\eta^4I^2\sum\limits_{l=1}^{L_c}G_l^2\\&\quad+2\beta\eta^2\sum_{m=1}^M\kappa_m\left((q_{s,m}-\bar{q}_s)^2+\bar{q}_s^2\right)\Lambda_m^2\\&\quad+\eta  \left(\sum\limits_{m=1}^M \Lambda_m^2 +4M\beta^2\eta^2I^2\sum\limits_{l=1}^{L_c}G_l^2\right)\\
&\quad- \frac{\eta}{2} \mathbb{E}\left[\left\|\sum\limits_{m=1}^M \varphi_m \nabla \mathcal{L}\left(\boldsymbol{w}^{t-1}\right)\right\|^2\right]\\&
\quad +4\beta\eta^2\mathbb{E}\left[\|\nabla\mathcal{L}(\boldsymbol{w}^{t-1})\|^2\right].
\end{aligned}
\label{fuluD}
\end{equation}

Next, we rearrange Eq. \eqref{fuluD} and divide its both sides by $\frac{\eta\left(\sum\limits_{m=1}^M\varphi_m\right)^2-8\beta\eta^2}{2}$:
\begin{equation}
\begin{aligned}
&\mathbb{E}\left[\|\nabla\mathcal{L}(\boldsymbol{w}^{t-1})\|^2\right]\leq\frac{2}{\eta\left(\sum\limits_{m=1}^M\varphi_m\right)^2-8\beta\eta^2}\cdot\\&\quad\Bigg(\mathbb{E}\left[\mathcal{L}\left(\boldsymbol{w}^{t-1}\right)\right]-\mathbb{E}\left[\mathcal{L}\left(\boldsymbol{w}^{t}\right)\right]+\frac{\sum\limits_{l=1}^{L} \beta\eta^2\sigma_l^2}{Mb}\\& \quad+\frac{2\beta\eta^2}{M^2} \sum\limits_{m=1}^M \Lambda_m^2+16\beta^3\eta^4I^2\sum\limits_{l=1}^{L_c}G_l^2\\&\quad+2\beta\eta^2\sum_{m=1}^M\kappa_m\left((q_{s,m}-\bar{q}_s)^2+\bar{q}_s^2\right)\Lambda_m^2\\&\quad+  \sum\limits_{m=1}^M \eta\Lambda_m^2 +4M\beta^2\eta^3I^2\sum\limits_{l=1}^{L_c}G_l^2\Bigg).
\end{aligned}
\end{equation}

For ensuring the convergence of split federated learning process, we let $\eta\left(\sum\limits_{m=1}^M\varphi_m\right)^2-8\beta\eta^2 > 0$, which leads to $0<\eta<\frac{\sum\limits_{m=1}^M\varphi_m}{8\beta}$. Furthermore, adding up the aforementioned terms from $t=1$ to $N$ and then dividing both sides by $N$ results in
\begin{equation}
\begin{aligned}
&\frac{1}{N}\sum\limits_{t=1}^N\mathbb{E}\left[\|\nabla\mathcal{L}(\boldsymbol{w}^{t-1})\|^2\right]\leq\frac{2}{\eta\left(\sum\limits_{m=1}^M\varphi_m\right)^2-8\beta\eta^2}\cdot\\&\quad\Bigg(\frac{\mathbb{E}\left[\mathcal{L}\left(\boldsymbol{w}^{0}\right)\right]-\mathbb{E}\left[\mathcal{L}\left(\boldsymbol{w}^{*}\right)\right]}{N}+\frac{\sum\limits_{l=1}^{L} \beta\eta^2\sigma_l^2}{Mb}\\& \quad+\frac{2\beta\eta^2}{M^2} \sum\limits_{m=1}^M \Lambda_m^2+16\beta^3\eta^4I^2\sum\limits_{l=1}^{L_c}G_l^2\\&\quad+2\beta\eta^2\sum_{m=1}^M\kappa_m\left((q_{s,m}-\bar{q}_s)^2+\bar{q}_s^2\right)\Lambda_m^2\\&\quad+ \sum\limits_{m=1}^M \eta\Lambda_m^2 +4M\beta^2\eta^3I^2\sum\limits_{l=1}^{L_c}G_l^2\Bigg),
\end{aligned}
\end{equation}
where $\boldsymbol{w}^*$ is the optimal model.

\section*{APPENDIX E}
We assume that each UAV has an identical successful sensing probability $q_{s,m}=q_s, \forall m \in \mathcal{M}$, thus we can have
\begin{equation}
\begin{aligned}
&\mathbb{E}\left[\sum\limits_{l=1}^M\sum\limits_{\substack{|\mathcal{M}_1|=l\\|\mathcal{M}_2|=M-l}}^{\mathcal{M}_1\cup \mathcal{M}_2=\mathcal{M}}\frac{1}{M}\sum\limits_{m_1 \in \mathcal{M}_1}\mathbf{g}_{m_1}^{t-1} \right.  \\
& \left.\quad\cdot\frac{\prod\limits_{m_1 \in \mathcal{M}_1}q_{s,m_1}\prod\limits_{m_2 \in \mathcal{M}_2}(1-q_{s,m_2})}{1-\prod\limits_{m \in M}(1-q_{s,m})}\right]\\&
=\frac{1}{M}\mathbb{E}\left[\sum\limits_{m=1}^M\frac{q_s^m(1-q_s)^{M-m}}{1-(1-q_s)^M}\sum\limits_{\substack{|\mathcal{M}_1|=m\\|\mathcal{M}_2|=M-m}}^{\mathcal{M}_1 \cup \mathcal{M}_2=\mathcal{M}}\sum\limits_{m_1 \in \mathcal{M}_1}\mathbf{g}_{m_1}^{t-1}\right]\\&
=\frac{1}{M}\mathbb{E}\left[\sum\limits_{m=1}^M\frac{q_s^m(1-q_s)^{M-m}}{1-(1-q_s)^M}C_{M-1}^{m-1}\sum\limits_{m \in \mathcal{M}}\mathbf{g}_{m}^{t-1}\right]
\\&
=\frac{q_s}{\left[1-(1-q_s)^M\right] \cdot M} 
\mathbb{E}\left[\sum\limits_{m=0}^{M-1}q_s^{m}(1-q_s)^{M-m-1} \right. \\
& \quad \left. C_{M-1}^{m} \sum\limits_{m \in \mathcal{M}} \mathbf{g}_{m}^{t-1} \right]\\&
=\frac{q_s}{\left[1-(1-q_s)^M\right] \cdot M}\sum\limits_{m\in \mathcal{M}}\mathbb{E}\left[\mathbf{g}_m^{t-1}\right].
\end{aligned}
\end{equation}
\bibliographystyle{IEEEtran}
\bibliography{ref}

@ARTICLE{10063977,
  author={Li, Xingyu and Qu, Zhe and Tang, Bo and Lu, Zhuo},
  journal={IEEE Transactions on Cybernetics}, 
  title={FedLGA: Toward System-Heterogeneity of Federated Learning via Local Gradient Approximation}, 
  year={2024},
  volume={54},
  number={1},
  pages={401-414},
  doi={10.1109/TCYB.2023.3247365}}

@ARTICLE{10438010,
  author={Huo, Xin and Zhang, Hao and Wang, Zhuping and Yan, Huaicheng and Liu, Chun},
  journal={IEEE Transactions on Cybernetics}, 
  title={An Efficient Matching Game Approach to Association Formation in {UAV}-Enabled Hierarchical Distributed Learning}, 
  year={2024},
  volume={54},
  number={10},
  pages={5696-5707},
  doi={10.1109/TCYB.2024.3357848}}

@INPROCEEDINGS{HouSFL,
  author={Hou, Xiangwang and Wang, Jingjing and Wang, Jiacheng and Du, Jun and Zhang, Zekai and Jiang, Chunxiao and Ren, Yong},
  booktitle={2025 IEEE International Conference on Communications (ICC), Montreal, QC, Canada}, 
  title={Energy-Efficient Federated Semi-Supervised Learning for {UAV}-Enabled Integrated Sensing, Computation, and Communication}, 
  year={2025},
  volume={},
  number={},
  pages={6680-6686},
  doi={10.1109/ICC52391.2025.11161423}}

@ARTICLE{wxh,
  author={Wang, Xianghe and Song, Shaoyang and Zhang, Zekai and Hou, Xiangwang and Li, Zhiying and Xing, Tianyu and Zhang, Xiao-Ping},
  journal={IEEE Transactions on Consumer Electronics}, 
  title={Split Federated Learning for Resource-Constrained Edge Computing Networks}, 
  year={2025},
  volume={71},
  number={4},
  pages={11001-11013},
  doi={10.1109/TCE.2025.3626460}}

@article{FMCW,
  title={Radar-based human-motion recognition with deep learning: Promising applications for indoor monitoring},
  author={Gurbuz, Sevgi Zubeyde and Amin, Moeness G},
  journal={IEEE Signal Processing Magazine},
  volume={36},
  number={4},
  pages={16--28},
  year={2019},
  publisher={IEEE}
}

@article{body,
  title={Simulation and analysis of human micro-Dopplers in through-wall environments},
  author={Ram, Shobha Sundar and Christianson, Craig and Kim, Youngwook and Ling, Hao},
  journal={IEEE Transactions on Geoscience and Remote Sensing},
  volume={48},
  number={4},
  pages={2015--2023},
  year={2010},
  publisher={IEEE}
}

@ARTICLE{ISAC-2306,
  author={Tang, Yao and Zhu, Guangxu and Xu, Wei and Cheung, Man Hon and Lok, Tat-Ming and Cui, Shuguang},
  journal={IEEE Transactions on Wireless Communications}, 
  title={Integrated Sensing, Computation, and Communication for UAV-assisted Federated Edge Learning}, 
  year={2025},
  volume={},
  number={},
  pages={1-1},
  doi={10.1109/TWC.2024.3523381}}

@inproceedings{A4,
author = {Krizhevsky, Alex and Sutskever, Ilya and Hinton, Geoffrey E.},
title = {ImageNet classification with deep convolutional neural networks},
year = {2012},
publisher = {Curran Associates Inc.},
booktitle = {2012 International Conference on Neural Information Processing Systems (NIPS), Lake Tahoe, Nevada},
pages = {1097–1105},
numpages = {9}
}

@inproceedings{A5,
  author       = {Karen Simonyan and
                  Andrew Zisserman},
  title        = {Very Deep Convolutional Networks for Large-Scale Image Recognition},
  booktitle    = {2015 International Conference on Learning Representations, (ICLR), San Diego, CA, USA},
  pages = {1-14, 2015}
}

@inproceedings{B1,
  title={Splitfed: When federated learning meets split learning},
  author={Thapa, Chandra and Arachchige, Pathum Chamikara Mahawaga and Camtepe, Seyit and Sun, Lichao},
  booktitle={2022 Association for the Advancement of Artificial Intelligence (AAAI), Vancouver, Canada},
  volume={36},
  number={8},
  pages={8485--8493, 2022}
}

@ARTICLE{B3,
  author={Yang, Ziyuan and Chen, Yingyu and Huangfu, Huijie and Ran, Maosong and Wang, Hui and Li, Xiaoxiao and Zhang, Yi},
  journal={IEEE Journal of Biomedical and Health Informatics}, 
  title={Dynamic Corrected Split Federated Learning With Homomorphic Encryption for {U}-Shaped Medical Image Networks}, 
  year={2023},
  volume={27},
  number={12},
  pages={5946-5957},
  doi={10.1109/JBHI.2023.3317632}}

@ARTICLE{B6,
  author={Xu, Ce and Li, Jinxuan and Liu, Yuan and Ling, Yushi and Wen, Miaowen},
  journal={IEEE Transactions on Wireless Communications}, 
  title={Accelerating Split Federated Learning Over Wireless Communication Networks}, 
  year={2024},
  volume={23},
  number={6},
  pages={5587-5599},
  doi={10.1109/TWC.2023.3327372}}

@ARTICLE{B7,
  author={Solat, Faranaksadat and Lee, Joohyung and Niyato, Dusit},
  journal={IEEE Wireless Communications Letters}, 
  title={Split Federated Learning-Empowered Energy-Efficient Mobile Traffic Prediction Over UAVs}, 
  year={2024},
  volume={13},
  number={11},
  pages={3064-3068},
  doi={10.1109/LWC.2024.3440397}}

@ARTICLE{B9,
  author={Wu, Wen and Li, Mushu and Qu, Kaige and Zhou, Conghao and Shen, Xuemin and Zhuang, Weihua and Li, Xu and Shi, Weisen},
  journal={IEEE Journal on Selected Areas in Communications}, 
  title={Split Learning Over Wireless Networks: Parallel Design and Resource Management}, 
  year={2023},
  volume={41},
  number={4},
  pages={1051-1066},
  doi={10.1109/JSAC.2023.3242704}}

@article{B10,
  title={AdaptSFL: Adaptive Split Federated Learning in Resource-constrained Edge Networks}, 
  author={Zheng Lin and Guanqiao Qu and Wei Wei and Xianhao Chen and Kin K. Leung},
  journal={ArXiv preprint ArXiv:2403.13101v3},
  year={2024}
}

@ARTICLE{B11,
  author={Wang, Shiqiang and Tuor, Tiffany and Salonidis, Theodoros and Leung, Kin K. and Makaya, Christian and He, Ting and Chan, Kevin},
  journal={IEEE Journal on Selected Areas in Communications}, 
  title={Adaptive Federated Learning in Resource Constrained Edge Computing Systems}, 
  year={2019},
  volume={37},
  number={6},
  pages={1205-1221},
  doi={10.1109/JSAC.2019.2904348}}

@ARTICLE{B12,
  author={Zhang, Junhe and Ni, Wanli and Wang, Dongyu},
  journal={IEEE Transactions on Vehicular Technology}, 
  title={Federated Split Learning with Model Pruning and Gradient Quantization in Wireless Networks, Early Access, {DOI}=10.1109/TVT.2024.3515083}, 
  year={2024},
  volume={},
  number={},
  pages={1-6},
  doi={10.1109/TVT.2024.3515083}}

@INPROCEEDINGS{B13,
  author={Shi, Wenqi and Zhou, Sheng and Niu, Zhisheng},
  booktitle={2020 IEEE International Conference on Communications (ICC), Dublin, Ireland}, 
  title={Device Scheduling with Fast Convergence for Wireless Federated Learning}, 
  volume={},
  number={},
  pages={1-6, 2020},
  doi={10.1109/ICC40277.2020.9149138}}

@ARTICLE{B14,
  author={Poposka, Marija and Pejoski, Slavche and Rakovic, Valentin and Denkovski, Daniel and Gjoreski, Hristijan and Hadzi-Velkov, Zoran},
  journal={IEEE Communications Letters}, 
  title={Delay Minimization of Federated Learning Over Wireless Powered Communication Networks}, 
  year={2024},
  volume={28},
  number={1},
  pages={108-112},
  doi={10.1109/LCOMM.2023.3337320}}

@INPROCEEDINGS{B15,
  author={Tran, Nguyen H. and Bao, Wei and Zomaya, Albert and Nguyen, Minh N. H. and Hong, Choong Seon},
  booktitle={2019 IEEE Conference on Computer Communications (INFOCOM), Paris, France}, 
  title={Federated Learning over Wireless Networks: Optimization Model Design and Analysis},
  volume={},
  number={},
  pages={1387-1395, 2019},
  doi={10.1109/INFOCOM.2019.8737464}}

@ARTICLE{9468714,
  author={Xie, Lifeng and Cao, Xiaowen and Xu, Jie and Zhang, Rui},
  journal={IEEE Transactions on Green Communications and Networking}, 
  title={{UAV}-Enabled Wireless Power Transfer: A Tutorial Overview}, 
  year={2021},
  volume={5},
  number={4},
  pages={2042-2064},
  doi={10.1109/TGCN.2021.3093718}}

@article{Zhang2024WhenUM,
  title={When {UAV} Meets Federated Learning: Latency Minimization via Joint Trajectory Design and Resource Allocation},
  author={Xuhui Zhang and Wenchao Liu and Jinke Ren and Huijun Xing and Gui Gui and Yanyan Shen and Shuguang Cui},
  journal={ArXiv preprint ArXiv:2412.07428},
  year={2024}
}

@ARTICLE{9220170,
  author={Tak, Afaf and Cherkaoui, Soumaya},
  journal={IEEE Network}, 
  title={Federated Edge Learning: Design Issues and Challenges}, 
  year={2021},
  volume={35},
  number={2},
  pages={252-258},
  doi={10.1109/MNET.011.2000478}}

@ARTICLE{10630700,
  author={Solat, Faranaksadat and Lee, Joohyung and Niyato, Dusit},
  journal={IEEE Wireless Communications Letters}, 
  title={Split Federated Learning-Empowered Energy-Efficient Mobile Traffic Prediction Over UAVs}, 
  year={2024},
  volume={13},
  number={11},
  pages={3064-3068},
  doi={10.1109/LWC.2024.3440397}}

@ARTICLE{9292475,
  author={Ng, Jer Shyuan and Lim, Wei Yang Bryan and Dai, Hong-Ning and Xiong, Zehui and Huang, Jianqiang and Niyato, Dusit and Hua, Xian-Sheng and Leung, Cyril and Miao, Chunyan},
  journal={IEEE Transactions on Intelligent Transportation Systems}, 
  title={Joint Auction-Coalition Formation Framework for Communication-Efficient Federated Learning in {UAV}-Enabled Internet of Vehicles}, 
  year={2021},
  volume={22},
  number={4},
  pages={2326-2344},
  doi={10.1109/TITS.2020.3041345}}

@INPROCEEDINGS{9593198,
  author={Li, Guoliang and Wang, Shuai and Li, Jie and Wang, Rui and Peng, Xiaohui and Han, Tony Xiao},
  booktitle={2021 IEEE International Workshop on Signal Processing Advances in Wireless Communications (SPAWC), Lucca, Italy}, 
  title={Wireless Sensing With Deep Spectrogram Network and Primitive Based Autoregressive Hybrid Channel Model}, 
  volume={},
  number={},
  pages={481-485, 2021},
  doi={10.1109/SPAWC51858.2021.9593198}}

@ARTICLE{6863654,
  author={Al-Hourani, Akram and Kandeepan, Sithamparanathan and Lardner, Simon},
  journal={IEEE Wireless Communications Letters}, 
  title={Optimal LAP Altitude for Maximum Coverage}, 
  year={2014},
  volume={3},
  number={6},
  pages={569-572},
  doi={10.1109/LWC.2014.2342736}}
\end{sloppypar}
\end{document}